\documentclass{ws-p9-75x6-50}
\usepackage{overcite}

\begin{document}

\title{Black holes and gamma ray bursts:\\ background for the theoretical model}

\author{Remo J. Ruffini}

\address{I.C.R.A. -- International Center for Relativistic Astrophysics
and
Physics Department, University of Rome ``La Sapienza", I-00185 Rome, Italy\\E-mail ruffini@icra.it}

\maketitle

\abstracts{
The idea that the vacuum polarization process occurring during gravitational collapse 
to a black hole endowed with electromagnetic structure (EMBH) could be the origin of 
gamma ray bursts (GRBs) is further developed. EMBHs in the range
3.2 -- 10$^6$ solar masses are considered. The formation of such an EMBH, 
the extraction of its mass-energy by reversible
transformations and the expansion of the pair-electromagnetic pulse (PEM pulse) 
are all examined within general relativity.
The PEM pulse is shown to accelerate particles to speeds with Lorentz gamma factors 
way beyond any existing experiment on Earth. 
Details of the expected burst structures and other observable 
properties are examined.
}

The Kerr-Newman mathematical solution of the Einstein-Maxwell equations, whose metric is \cite{nccept65}
\begin{equation}
ds^2=\Sigma\Delta^{-1}dr^2+\Sigma d\vartheta^2+\Sigma^{-1}\sin^2\vartheta\left[adt-\left(r^2+a^2\right)d\varphi\right]^2-\Sigma^{-1}\Delta\left[dt-a\sin^2\vartheta d\varphi\right]^2 ,
\label{KNmetric}
\end{equation}
where
\begin{equation}
\Delta=r^2-2Mr+Q^2+a^2\, ,\quad \Sigma=r^2+a^2\cos^2\vartheta\, ,
\label{KNmetric2}
\end{equation}
has been one of the most powerful theoretical tools for probing spacetime structure by relating the concepts of mass $M$, charge $Q$ and angular momentum $L$, with $a=L/M$, within a relativistic field theory. This quantities have been expressed in geometrical units (see e.g. Ruffini in Ref.~\citen{gr78}). The aim of this talk is to show how this metric is currently becoming the fundamental theoretical tool for explaining the most energetic events ever discovered in our universe: the gamma ray bursts (GRBS). Essential for doing this is to consider the quantum vacuum polarization process in a Kerr-Newmann geometry and relate it to the reversible transformations of the black hole and to its  mass-energy formula. This leads us to believe that in these events we are witnessing energy extraction from a black hole for the first time. We will take a somewhat historical approach to recover the ``crescendo" of this field of research and recollect some of the most crucial moments which are making this discovery possible.

\section{Early steps in the study of black holes}

With the discovery of pulsars and of the pulsar in the Crab Nebula in particular, the year 1968 can be considered the birth date of relativistic astrophysics. The observation of the period and the slow-down rate of the Crab pulsar unequivocally identified the first neutron star in the galaxy and showed that the energy source of pulsars is just the rotational energy of a neutron star.

\begin{figure}
\vspace{-.5cm}
\epsfxsize=6.0cm
\begin{center}
\mbox{\epsfbox{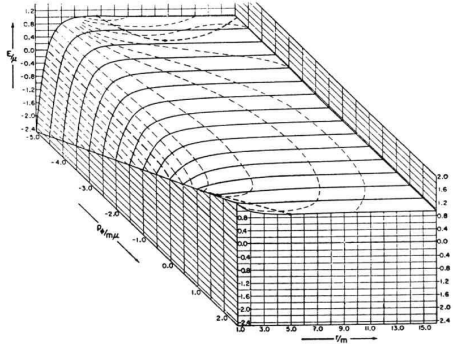}}
\end{center}
\vspace{-0.2cm}
\caption[]{``Effective potential'' around a Kerr black hole, see Ruffini and Wheeler 1971.}
\label{poten}
\end{figure}

At the time I was in Princeton as a  postdoctoral fellow in the group of John A. Wheeler, then as a member of the Institute for Advanced Study, and later as an assistant professor at the university. The excitement over the neutron star discovery  boldly led us to explore the paper by Robert Oppenheimer and Snyder ``on continued gravitational contraction''\cite{snyder}: this opened up a new field of research to which I have dedicated the remainder of my life and it is still producing important results today. An ``effective potential'' technique had been used  by Carl St\o rmer in the 1930s in studying the trajectories of cosmic rays in the Earth's magnetic field (St\o rmer 1934)\cite{s34}. In the fall of 1967 Brandon Carter visited Princeton and presented his remarkable mathematical work leading to the separability of the Hamilton-Jacobi equations for the trajectories of charged particles in the field of a Kerr-Newmann geometry (Carter 1968)\cite{bc}. This visit had a profound impact on our small group. It was Johnny Wheeler's idea to exploit the analogy between the trajectories of cosmic rays and the spacetime trajectories of charged particles in general relativity, using the St\o rmer ``effective potential'' technique in order to obtain physical consequences from Carter's set of first order differential equations. I remember the excitement of preparing the $2m\times 2m$ grid plot of the effective potential for particles around a Kerr black hole which finally appeared later in print (Rees, Ruffini and Wheeler 1973, 1974\cite{rrw}); see Fig.~(\ref{poten}). 

\begin{figure}
\vspace{-.5cm}
\epsfxsize=6.0cm
\begin{center}
\mbox{\epsfbox{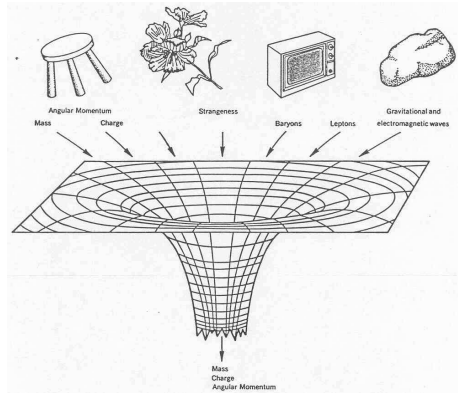}}
\end{center}
\vspace{-0.2cm}
\caption[]{The black hole uniqueness theorem.}
\label{tvsetbh}
\end{figure}

Out of this work came the celebrated result for the maximum binding energy  $1 - {1 \over \sqrt{3}}\sim42\%$ for corotating orbits and $1-{5\over 3\sqrt{3}}\sim 3.78\%$ for counter-rotating orbits in the Kerr geometry. We were pleased to be later associated with Brandon Carter in a ``gold medal'' award for this work presented by Yevgeny Lifshitz: in the fourth and last edition of volume 2 of the Landau and Lifshitz series ({\it The Classical Theory of Fields\/}, 1975), both Brandon's work and my own work with Wheeler were proposed as named exercises for bright students! 
In the article ``Introducing the Black Hole" (Ruffini and Wheeler 1971)\cite{rw71} we proposed the famous ``uniqueness theorem'' stating that black holes can only be characterized by their mass-energy $E$, charge $Q$ and angular momentum $L$. This analogy between a black hole and a very elementary physical system  was imaginatively represented by Johnny in a  very unconventional figure in which TV sets, bread, flowers and other objects lose their characteristic features and merge in the process of gravitational collapse into the three fundamental parameters of a black hole, see Fig.~\ref{tvsetbh}. That picture became the object of a great deal of lighthearted discussion in the physics community. A proof of this uniqueness theorem, satisfactory for some cases of astrophysical interest, has been obtained after twenty five years of meticulous mathematical work (see e.g., Regge and Wheeler\cite{ReggeW}, Zerilli\cite{Zerilli1,Zerilli2}, Teukolsky\cite{teukolsky}, C.H. Lee\cite{lee}, Chandrasekhar\cite{chandra}). However, a rigorous proof  still presents some outstanding technical difficulties in its most general form. Possibly some progress will be reached in the near future with the help of computer algebraic manipulation techniques to overcome the extremely difficult mathematical calculations (see e.g., Cruciani (1999),\cite{cru} Cherubini and Ruffini  (2000),\cite{chrr} Bini et al.\ (2001),\cite{bcjr1} Bini et al.\  (2001)\cite{ba02}).

This ansatz, which at first appeared to be almost trivial, has shown itself to be one of the most difficult to be proved, unsurpassed in difficulty both in mathematical physics and relativistic field theories. I am convinced that the mathematical tools developed in order to prove this uniqueness theorem will have profound implications for understanding fundamental laws of physics. 

\section{From ``dead'' to ``live'' black holes}

We were still under the sobering effects of the pulsar discovery and the explanation by T. Gold and A. Finzi that the rotational energy of the neutron star had to be the energy source of pulsars when the first meeting of the European Physical Society took place in Florence in 1969. There Roger Penrose\cite{p69} advanced the possibility that, much like in the case of pulsars, the rotational energy of black holes could also be extracted in principle.

\begin{figure}
\vspace{-.5cm}
\epsfxsize=6.0cm
\begin{center}
\mbox{\epsfbox{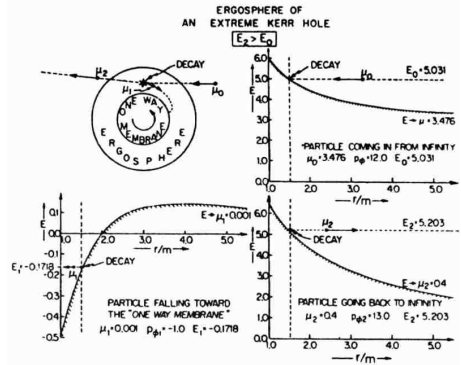}}
\end{center}
\vspace{-0.2cm}
\caption[]{Decay of a particle of rest-plus-kinetic energy $E_\circ$ into a particle which is captured by the black hole with positive energy as judged locally, but negative energy $E_1$ as judged from infinity, together with a particle of rest-plus-kinetic energy $E_2>E_\circ$ which escapes to infinity. The cross-hatched curves give the effective potential (gravitational plus centrifugal) defined by the solution $E$ of Eq.(2) for constant values of $p_\phi$ and $\mu$. (Figure and caption reproduced from Christodoulou 1970\cite{chris1}, in turn reproduced before its original publication in Ref.~\citen{rw71} with the kind permission of Ruffini and Wheeler.)}
\label{pic1}
\end{figure}

The first specific example of such an energy extraction process by a gedanken experiment was given using the above-mentioned effective potential technique in Ruffini and Wheeler (1970)\cite{ruffx}, see Figure (\ref{pic1}), and then later by Floyd and Penrose (1971)\cite{fr}. The reason for showing this figure here is threefold: a) to recall the first explicit computation and b) to recall the introduction of the ``ergosphere'', the region between the horizon of a Kerr-Newmann metric and the surface of infinite redshift were the energy extraction process can occur, and also c) to emphasize how contrived but also conceptually novel such an energy-extraction mechanism can be. It is a phenomenon which is not localized at a point but which occurs in an entire region: a global effect which relies essentially on the concept of a field. It only works for very special parameters and is in general associated with a reduction of the rest mass of the particle involved in the process. It is almost trivial to slow down the rotation of a black hole and increase its horizon size by  accretion of counter-rotating particles, but it is extremely difficult to extract the rotational energy from a black hole by a slow-down process, as pointed out by the example in Fig.~(\ref{pic1}). The establishment of this analogy offered us the opportunity to appreciate even more the profound difference between seemingly similar effects in general relativity and classical field theories. In addition to the existence of totally new phenomena, like the dragging of inertial frames around a rotating black hole for example, we had the first glimpse of an entirely new field of theoretical physics present in and implied by the field equations of general relativity. The deep discussions of these problems with Demetrios Christodoulou, who was a 17 year old Princeton student at the time,  my first graduate student, led us to the discovery of  the existence in black hole physics of the ``reversible and irreversible transformations." 

\begin{figure}
\vspace{-.5cm}
\epsfxsize=6.0cm
\begin{center}
\mbox{\epsfbox{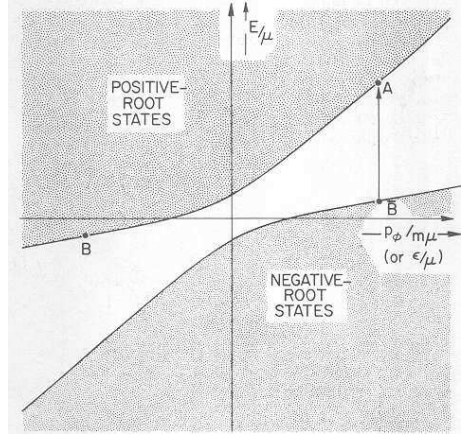}}
\end{center}
\vspace{-0.2cm}
\caption[]{Reversing the effect of having added to the black hole one particle (A) by adding another particle (B) of the same rest mass but opposite angular momentum and charge in a ``positive-root negative-energy state''. Addition of B is equivalent to subtraction of $B^-$. Thus the combined effect of the capture of particles A and B is an increase in the mass of the black hole given by the vector $B^-A$. This vector vanishes and reversibility is achieved when and only when the separation between positive root states and negative root states is zero, in which case the hyperbolas coalesce to a straight line. Reproduced from Christodoulou and Ruffini \cite{ruffc}.}
\label{posneg}
\end{figure}

It was in fact by analyzing the capture of test particles by a black hole endowed with electromagnetic structure, for short an EMBH,  that we identified a set of limiting transformations which did not affect the surface area of an EMBH. These special transformations had to be performed very slowly, with a limiting value of zero kinetic energy on the horizon of the EMBH, see Fig.~\ref{posneg}. It became clear that the total energy of an EMBH could in principle be expressed as a ``rest energy,'' a ``Coulomb energy'' and a ``rotational energy.'' The rest energy is ``irreducible'', the other two being submitted to positive and negative variations, corresponding respectively to processes of addition and subtraction, namely extraction, of energy.

While Wheeler was mainly attracted by the thermodynamics analogy, I addressed with Demetrios the issue of the energetics of EMBHs using the tools of reversible and irreversible transformations. We obtained the mass-energy formula for black holes (Christodoulou and Ruffini 1971)\cite{ruffc}: 
\begin{eqnarray}
E^2&=&M^2c^4=\left(M_{\rm ir}c^2 + {Q^2\over2\rho_+}\right)^2+{L^2c^2\over \rho_+^2}\,,\label{em}\\
S&=& 4\pi \rho_+^2=4\pi \left(r_+^2+{L^2\over c^2M^2}\right)=16\pi\left({G^2\over c^4}\right) M^2_{\rm ir}\,,
\label{sa}
\end{eqnarray}
with
\begin{equation}
{1\over \rho_+^4}\left({G^2\over c^8}\right)\left( Q^4+4L^2c^2\right)\leq 1\,,
\label{s1}
\end{equation}
where $M_{\rm ir}$ is the irreducible mass, $S$ is the horizon surface area, and extreme black holes satisfy the equality in eq.~(\ref{s1}). Here, for reasons which will become clear in the following, I express the mass-energy formula using $r_{+}$, the horizon radius, and $\rho_+$, the quasi-spheroidal cylindrical coordinate of the horizon evaluated at the equatorial plane, defined by Eq.~(\ref{sa}). For convenience here and in the following I use c.g.s units. Although the mass-energy formula has been obtained using test particles in a ``gedanken process,'' its validity is clearly general and does not depend on the particular ``gedanken process'' used for its determination.

The crucial point is that transformations at constant surface area of the black hole, namely reversible transformations, can release an energy up to 29\% of the mass-energy of an extremely rotating black hole and up to 50\% of the mass-energy of an extremely magnetized and charged black hole. Since my Les Houches lectures ``On the energetics of black holes'' (B.C. De Witt 1973)\cite{dw73}, I introduced the concepts of ``live'' black holes, endowed with mass-energy, rotation and angular momentum and ``dead'' black holes characterized by their masses alone: one of my main research goals since then has been to identify an astrophysical setting where the extractable mass-energy of the black hole could manifest itself. As we will see in the following, I propose that this extractable energy of an EMBH is the energy source of gamma ray bursts (GRBs).

\section{The paradigm for the identification of the first ``black hole'' in our galaxy and the development of X-ray astronomy.}

The launch of the ``Uhuru'' satellite by the group directed by R. Giacconi, dedicated to the first examination of the universe in X-rays,  marked a fundamental leap forward and generated a tremendous momentum in the field of relativistic astrophysics.  The very fortunate collaboration soon established with simultaneous observations in the optical and radio wavelengths generated high quality data on binary star systems composed of a normal star being stripped of matter by a compact massive companion star: either a neutron star or a black hole.

The ``maximum mass of a neutron star'' was the subject of the thesis of C. Rhoades, my second graduate student at Princeton. A criteria was found there to overcome fundamental unknowns about the behavior of matter at supranuclear densities by establishing an absolute upper limit to the neutron star mass based only on general relativity, causality and the behavior of matter at nuclear and subnuclear densities (Rhoades and Ruffini 1974)\cite{rr74}.

The three essential components in establishing the paradigm for the identification of the first black hole in Cygnus X1 (Leach and Ruffini 1973)\cite{lr73} were
\begin{itemize}
\item the ``black hole uniqueness theorem,'' implying the axial symmetry of the configuration and the absence of regular pulsations from black holes,
\item the ``effective potential technique,'' determining the efficiency of the energy emission in the accretion process, and 
\item  the ``upper limit on the maximum mass of a neutron star,'' discriminating between an unmagnetized neutron star and a black hole.
\end{itemize}
I also presented these results in a widely attended session chaired by Johnny at the 1972 Texas Symposium in New York, extensively reported on by the New York Times. The New York Academy of Sciences which hosted the symposium had just awarded me their Cressy Morrison Award for my work on neutron stars and black holes. Much to their dismay I never wrote the paper for the proceedings since it coincided with the one submitted for publication (Leach and Ruffini 1973)\cite{lr73}.

The definition of the paradigm did not come easily but slowly matured after innumerable discussions with R. Giacconi and H. Gursky and was finally summarized in the splendid occasion of my talk at the Sixteenth Solvay Conference on Astrophysics and Gravitaion held at the University of Bruxelles in September 1973\cite{rufsolv} and in two books: (Gursky and Ruffini 1975)\cite{gr75} and (Giacconi and Ruffini 1978)\cite{gr78}. These results were important in identifying the first black hole in our galaxy: Cygnus X1. The energy source of these binaries was also totally new: the energy released due to the increase of the numerical value of the gravitational binding energy of the matter accreting onto a neutron star or into a black hole. In this accretion process the black hole had a passive role just creating the deep potential well in order to release the observed X-ray flux, but not yet the active role connected with an energy extraction process from the black hole. All this paved the way to the goal I was constantly pursuing of identifying an astrophysical process to access the extractable energy of a black hole.

\section{The Heisenberg-Euler critical capacitor and vacuum polarization around a macroscopic black hole}

\begin{figure}
\vspace{-.5cm}
\epsfxsize=\hsize
\begin{center}
\mbox{\epsfbox{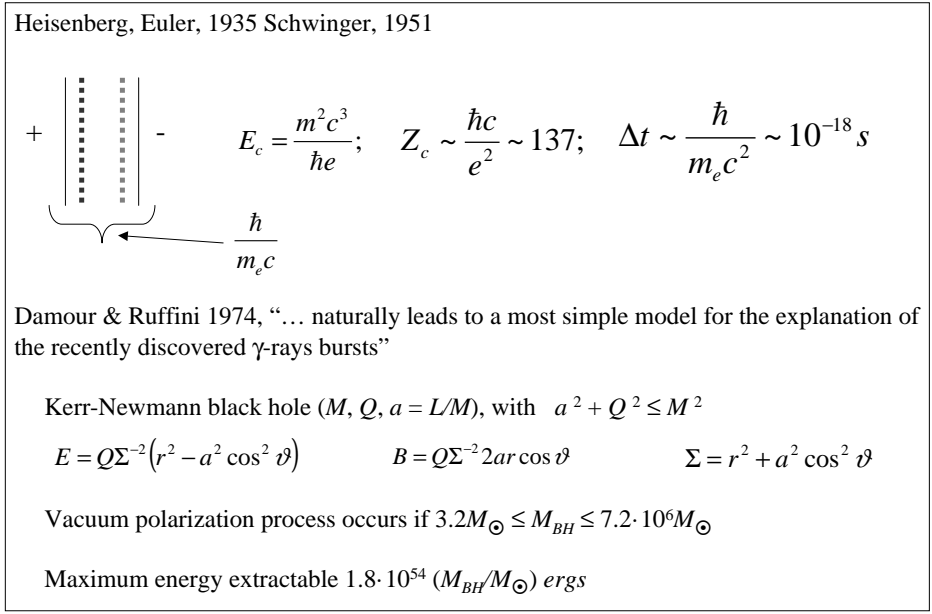}}
\end{center}
\vspace{-0.2cm}
\caption[]{Summary of the EMBH vacuum polarization process. See Damour \& Ruffini (1975)\cite{dr75} for details.}
\label{capdiap}
\end{figure}

In 1975, following the work on the energetics of black holes (Christodoulou and Ruffini 1971)\cite{ruffc}, we pointed out (Damour and Ruffini, 1975)\cite{dr75} the existence of the vacuum polarization process {\it a' la} Heisenberg-Euler-Schwinger (Heisenberg and Euler 1935\cite{he35}, Schwinger 1951\cite{s51}) around black holes endowed with electromagnetic structure (EMBHs). This work matured trough two very pleasant circumstances: 1) the leasurely discussions I had with Werner Heisenberg on this subject in M\"unich, Washington and Stanford, and 2) the coming to Princeton from Paris of Thibault Damour, with all his splendid mathematical craftmanship, in order to prepare his thesis under my supervision. With Thibault we proved that such vacuum polarization process can only occur for EMBHs of mass smaller then $7.2\cdot 10^{6}M_\odot$. The basic energetics implications were contained in Table~1 of that paper (Damour and Ruffini, 1975)\cite{dr75}, where it was also shown that this process is almost reversible in the sense introduced by Christodoulou and Ruffini (1971)\cite{ruffc} and that it extracts the mass energy of an EMBH very efficiently. We also pointed out that this vacuum polarization process around an EMBH offered a natural mechanism for explaining the GRBs, just discovered at the time. However, our mechanism had a most peculiar prediction: the characteristic energetics of the burst should be of  $\sim 10^{54}$ ergs, see Fig.~\ref{capdiap}. However, nothing at the time was known about either the distances or the energetics of GRBs. For details see Ruffini in Ref.~\citen{gr78}.

It is appropriate to remark that the process of vacuum polarization considered in Ref.~\cite{dr75} is profoundly different from the Hawking process. We recall that the Hawking radiation temperature $T$, time scale $\tau$ and flux $\Phi$ are given respectively by:
\begin{equation}
\begin{array}{c}
T\simeq 0.62\times 10^{-7} \frac{M_{\odot}}{M}\, {\rm K}\, ,\quad \tau\simeq\frac{E}{dE/dt}\simeq 2\times 10^{63}{\left(\frac{M}{M_{\odot}}\right)}^3\, {\rm years}\, ,\\
\Phi=\frac{dE}{dt}\simeq 10^{-22}{\left(\frac{M_{\odot}}{M}\right)}^2\, {\rm erg/s}\, .
\end{array}
\label{hawking1}
\end{equation}
For $M=10M_{\odot}$ we would obtain:
\begin{equation}
T\simeq 6.2\times 10^{-9}\, {\rm K}\, , \quad \tau\simeq 10^{66}\, {\rm years}\, , \quad \frac{dE}{dt}\simeq 10^{-24}\, {\rm erg/s}\, .
\label{hawking2}
\end{equation}

\section{Three ``happenings'' related to GRBs}

The case of gamma ray bursts is very intriguing for me personally. There have been moments in my life which appear to have been intertwined with some of the events that are leading to the understanding of such extremely unique phenomena. Each  scientific contribution I have made and even apparently occasional occurrences in my life, seemingly disconnected, appear to acquire a special meaning in reaching  an understanding of such phenomena. 

The first of such happenings was the collaboration with John Wheeler started at Princeton  in 1967  and with Yakov Borisovich Zel'dovich started in 1968 in Moscow. Both these scientists were the founders of research in relativistic astrophysics in their countries. But both of them played a key role in the nuclear arms race in the USA and Soviet Union respectively, before addressing their attention to the implications of Einstein's theory for astrophysics. Johnny Wheeler was a leader in the project which exploded the first American H-bomb. Ya.~B. Zeldovich, after having contributed to the defense of his country with the invention of the Katiuscia rockets, had developed the Soviet atomic and H-bombs with Andrej Sakahrov. He had also proposed a most unusual and repelling project: to have an H-bomb explode on the far side of the moon to demonstrate simultaneously the ``maturity'' of the nuclear and space technology reached by the Soviet Union in the early sixties. 

\begin{figure}
\vspace{-.5cm}
\epsfxsize=6.0cm
\begin{center}
\mbox{\epsfbox{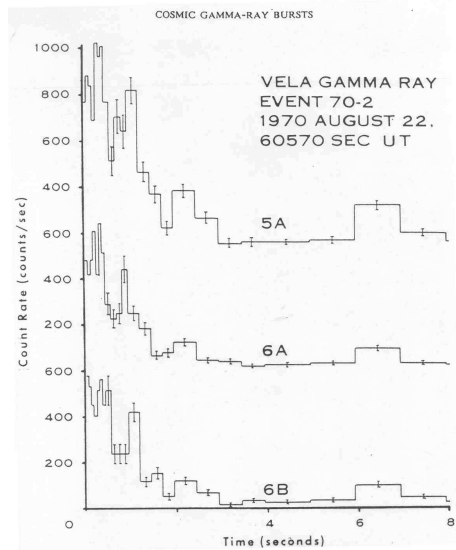}}
\end{center}
\vspace{-0.2cm}
\caption[]{One of the first GRBs observed by the Vela satellite. Reproduced from Strong in Gursky \& Ruffini (1975)\cite{gr75}.}
\label{velaburst}
\end{figure}

The second happening occurred in 1975. H. Gursky and myself had been invited by the AAAS to organize a session on neutron stars, black holes and binary X-ray sources for their annual meeting in San Francisco, Gursky \& Ruffini (1975)\cite{gr75}. During the preparation of the meeting we heard that some observations made by the military Vela satellites, developed to monitor the Limited Test Ban Treaty of 1963 banning atomic bomb explosions, had just been unclassified. Undoubtedly the  unorthodox proposal of Zel'dovich  had been among the motivations for developing such a grandiose military monitoring system. We asked Ian B. Strong to report on these just observed-released gamma ray bursts (GRBs) for the first time in a public meeting (Strong 1975)\cite{gr75}, see Fig.~\ref{velaburst}.

\begin{figure}
\vspace{-.5cm}
\epsfxsize=\hsize
\begin{center}
\mbox{\epsfbox{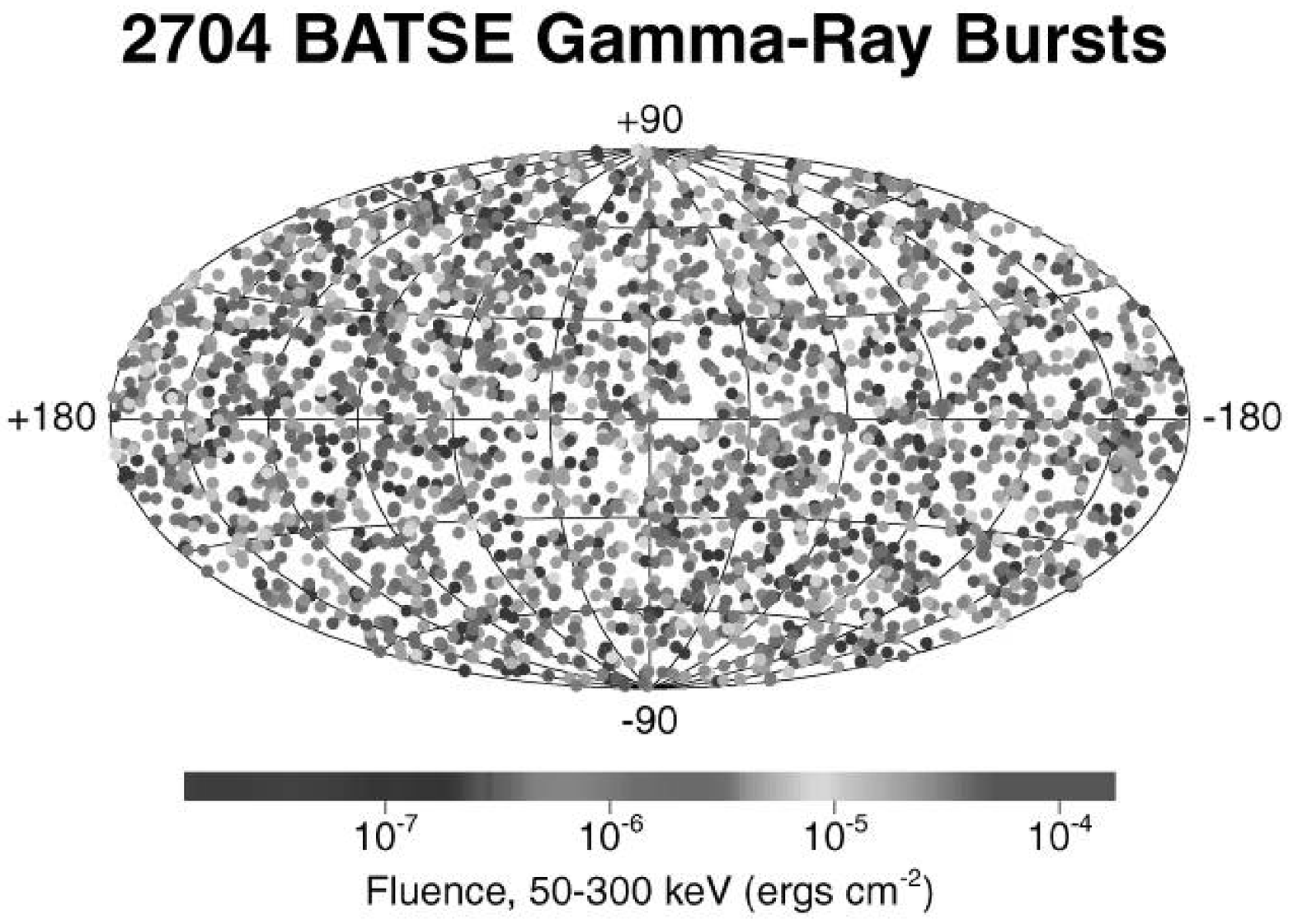}}
\end{center}
\vspace{-0.2cm}
\caption[]{Angular distribution of GRBs in galactic coordinates from the Compton GRO satellite.}
\label{batsedist}
\end{figure}

It was clear from the earliest observations that these signals were not coming either from the Earth or the planetary system. By 1991 a great improvement in knowledge of the distribution of the GRBs came with the NASA launch of the Compton Gamma Ray Observatory, which in ten years of observations gave beautiful evidence for the perfect isotropy of the angular distribution of the GRB sources in the sky, see Fig.~\ref{batsedist}. The sources had to either be at cosmological distances or very close to the solar system in order not to reflect the anisotropic galactic  distribution.

The third happening occurred in 1989 when I was elected president of the scientific committee of the Italian Space Agency (ASI) and the committee found itself involved with the scrutiny of the first Italian scientific satellite: the SAX satellite. The project was a collaboration between Italy and the Netherlands: The total estimated cost was roughly 50 million US dollars, fairly shared by the two partners: roughly 25 million from Italy and 25 million from the Netherlands. The satellite was supposed to fly in 1985. The program had already been delayed four years by the time our scrutiny started. The costs had correspondingly skyrocketed to almost 250 million US dollars, ``fairly" shared by the partners: roughly 225 million from Italy and 25 from the Netherlands... The real moment of panic in the ASI scientific committee came when we learned that the Dutch had run out of money! They could not afford to pay for the wide field X-ray cameras that they were supposed to contribute. The scientific committee decided to intervene offering to pay roughly six million US dollars from the limited budget of our committee in order to avoid any further delay and especially to avoid the loss  of one of the crucial instruments of the scientific mission. As the delays were augmenting and the expenses of the mission were correspondingly ``skyrocketing'' further, the ambiance soon deteriorated to inadmissible pressures. Before quitting ASI I insisted on the imperative to accomplish the mission no matter what. The satellite was finally launched in 1996 at a cost still unknown today. Months after the launch three of the four gyroscopes failed adding difficulties for an effective pointing capability.

In spite of all that, thanks also to the determined action of a small number of dedicated young physicists educated at ``La Sapienza'' (First University of Rome) who had joined the Milano-Palermo based original team, the newly named Beppo-SAX satellite was able to conclude one of the most successful scientific missions ever in astronomy and astrophysics.  By using the wide field X-ray cameras very effectively, they discovered the afterglows of the GRBs, which in turn have allowed the optical identification of the sources and the determination of their cosmological nature\cite{c97}, implying for the GRB sources an energetic typically of $10^{54} {\rm ergs/pulse}$. This is exactly the one which was the characteristic feature predicted in Damour and Ruffini (1975)\cite{dr75}. I am also happy to see that the completion of the wide field X-ray camera has been essential to the identification of the afterglow and that the many imperatives to conclude the mission have lead to a successful epilogue.

Thinking about this past situation with hindsight, I have reached a rather unorthodox conclusion: if SAX had flown on time in 1985 and possibly within its planned budget, it would have been a managerial success and quite a savings for the Italian treasury, but very likely not a scientific success. The reason is that in 1985  neither the Space Telescope nor the very large telescopes like KECK  and  VLT, which have been essential for the optical identification of the GRBs and the establishment of their cosmological distances, were functioning. Of course I do not want to make propaganda in favor of wrongdoing, but it appears as if a tremendous force directs human actions not only by exploiting great scientific ideas but by making use as well of not so effective management in order to reach an important final scientific goal!

\begin{figure}
\vspace{-.5cm}
\epsfysize=17.0cm
\begin{center}
\mbox{\epsfbox{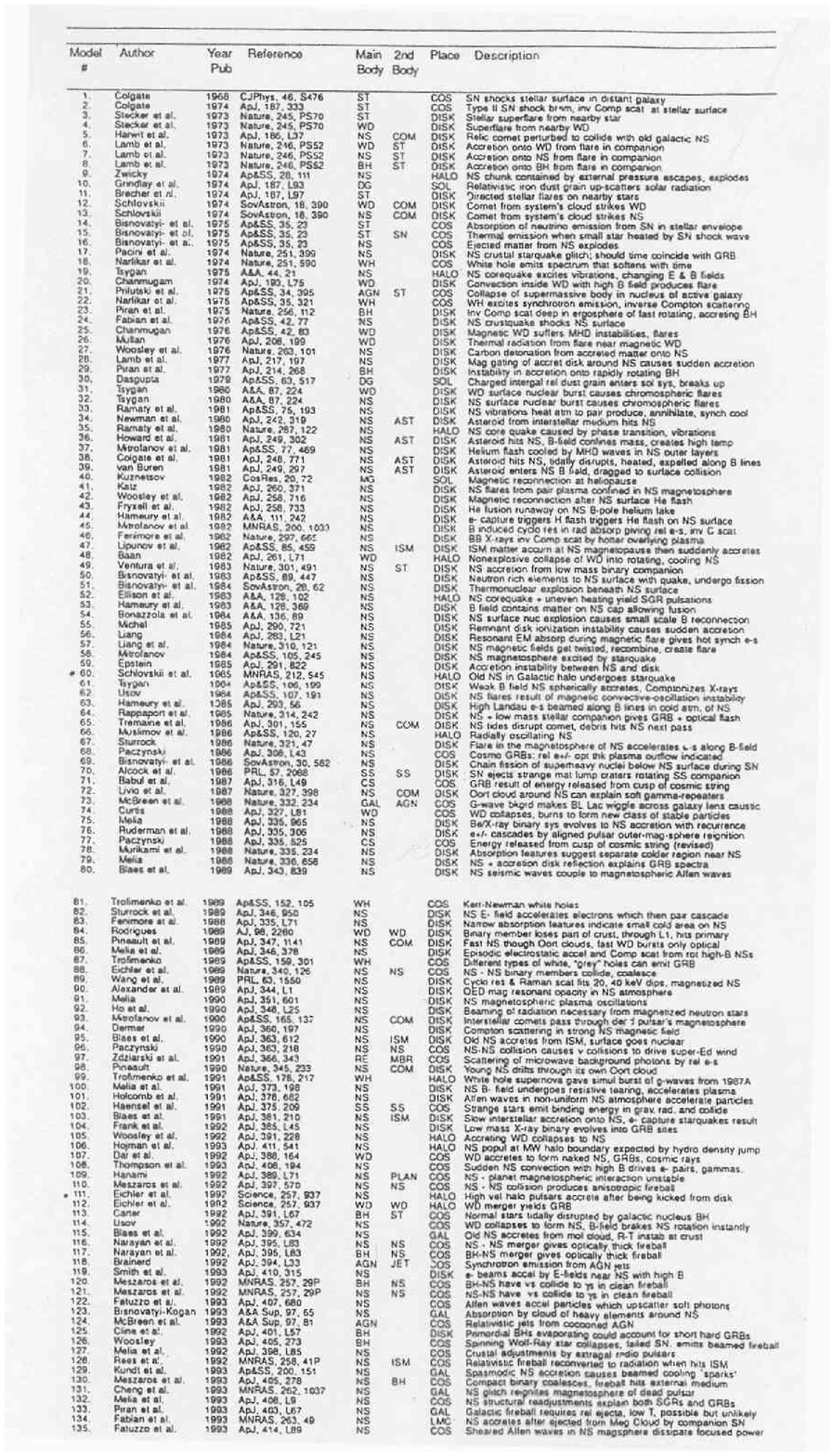}}
\end{center}
\vspace{-0.2cm}
\caption[]{Partial list of theories before the Beppo-SAX, from the talk presented at MGIXMM\cite{mgixmm}.}
\label{teofig}
\end{figure}

While the expenditures on Beppo-SAX were ``skyrocketing'', equally increasing exponentially were the numbers of competing theories trying  to explain GRBs. See a partial list in Fig.~\ref{teofig}. The observations of the Beppo-SAX satellite had a very sobering effect on the theoretical developments for GRB models. Almost all of the existing theories (see above partial list) were at once wiped out, not being able to fit the stringent energetics requirements imposed by the observations.

Particularly constrained by the observations were models based on the Hawking radiation process (see Tab.~\ref{tab1}): these show probably the largest discrepancies between a theoretical prediction and an observed phenomena ever recorded in human history.

\begin{table}
\caption[]{Hawking radiation process versus GRB observations.}
\footnotesize
\vspace{0.5cm}
For the correct value of the energetics $E_{tot} \simeq 10M_{\odot} \simeq 10^{55}$erg, we have:\\
\begin{center}
\begin{tabular}{|c|c|c|}
\hline
theoretical value & observed value & discrepancy \\
\hline
 & & \\
$T=6.2\cdot10^{-9}$\,K  &  $T \simeq 10^8$\,K  & $ \sim 10^{-17}$  \\
 & & \\
$\tau=10^{73}$\,sec  &  $\tau \simeq 1$\,sec & $\sim 10^{73}$ \\
 & & \\
$\left(\frac{dE}{dt}\right)=10^{-24}$\,erg/sec  
&  $\left(\frac{dE}{dt}\right) \simeq 10^{54}$\,erg/sec  &  $\sim 10^{-78}$\\
 & & \\
\hline
\end{tabular}
\end{center}
\vspace{0.5cm}
For the correct value of the time scale $\tau \simeq 1$ sec, we have:\\
\begin{center}
\begin{tabular}{|c|c|c|}
\hline
theoretical value & observed value & discrepancy \\
\hline
 & & \\
$E_{tot} \simeq 10^{-24} M_{\odot} \simeq 10^{30}$\,erg  &  $E_{tot} \simeq 10^{55}$\,erg & $\sim 10^{-25}$ \\
 & & \\
$T=10^{17}$\,K  &  $T \simeq 10^{8}$\,K & $\sim 10^{9}$ \\
 & & \\
$\left(\frac{dE}{dt}\right)=10^{26} $\,erg/sec  &  $\left(\frac{dE}{dt}\right) \simeq 10^{54} $\,erg/sec  &  $\sim 10^{-28}$\\
 & & \\
\hline
\end{tabular}
\end{center}
\vspace{0.5cm}
For the correct value of the spectrum energy $T=10^{8}$\,K, we have:\\
\begin{center}
\begin{tabular}{|c|c|c|}
\hline
theoretical value & observed value & discrepancy \\
\hline
 & & \\
$E_{tot} \simeq 10^{-9} M_{\odot} \simeq 10^{45}$\,erg  &  $E_{tot} \simeq 10^{55}$\,erg & $\sim 10^{-10}$ \\
 & & \\
$\tau=10^{43}$\,sec  &  $\tau \simeq 1$sec & $\sim 10^{43}$ \\
 & & \\
$\left(\frac{dE}{dt}\right)=10^{-4} $\,erg/sec  &  $\left(\frac{dE}{dt}\right) \simeq 10^{54} $\,erg/sec  &  $\sim 10^{-58}$\\
 & & \\
\hline
\end{tabular} 
\end{center}  
\label{tab1}
\end{table}

\section{A new paradigm for the EMBH formation}\label{newparadigm}

The enormous energy requirements of GRBs evidenced by the Beppo-SAX satellite, very similar to the ones predicted in Damour \& Ruffini (1975),\cite{dr75} convinced us to return to our earlier work in studying more accurately the process of vacuum polarization and the region of pair creation around an EMBH.

In our theoretical approach, we claim that through the observations of GRBs we are witnessing the formation of an EMBH and therefore are following the process of gravitational collapse in real time. Even more importantly, the tremendous energies involved in the energetics of these sources have their origin in the extractable energy of black holes given in Eqs.~(1)--(3) above.

Various models have been proposed in order to extract the rotational energy of black holes by processes of relativistic magnetohydrodynamics (see e.g., Ruffini and Wilson (1975)\cite{rw75}). It should be expected, however, that these processes are relevant over the long time scales characteristic of accretion processes.
In the present case of gamma ray bursts a sudden mechanism appears to be at work on time scales of the order of few seconds or shorter and they are naturally explained by the vacuum polarization process introduced in Damour \& Ruffini (1975)\cite{dr75}.

Although our major effort has been directed towards discovering the consequences of our model for the observed properties of GRBs starting from an already existing EMBH, some attention has been given to the issue of how an EMBH can in fact originate. Such a problem has been debated from many years since the earliest discussions in 1970 in Princeton.  Here I would like to propose a clarification and a change of paradigm.

All considerations on the electric charge of stars have been traditionally directed toward the presence of a net charge on the star surface in a steady state condition, from the classic work by Shvartsman\cite{s70} all the way to the fundamental book by Punsly.\cite{punsly_book}  The charge separation can occur in stars endowed with rotation and magnetic field and surrounded by plasma, as in the case of Goldreich and Julian (1969)\cite{gj69}, or in the case of absence of both magnetic field and rotation, the electrostatic processes can be related to the depth of the gravitational well, as in the treatment of Shvartsman\cite{s70}. However, in neither case is it possible to reach the condition of the overcritical field needed for pair creation.

\begin{figure}
\vspace{-.5cm}
\epsfxsize=\hsize
\begin{center}
\mbox{\epsfbox{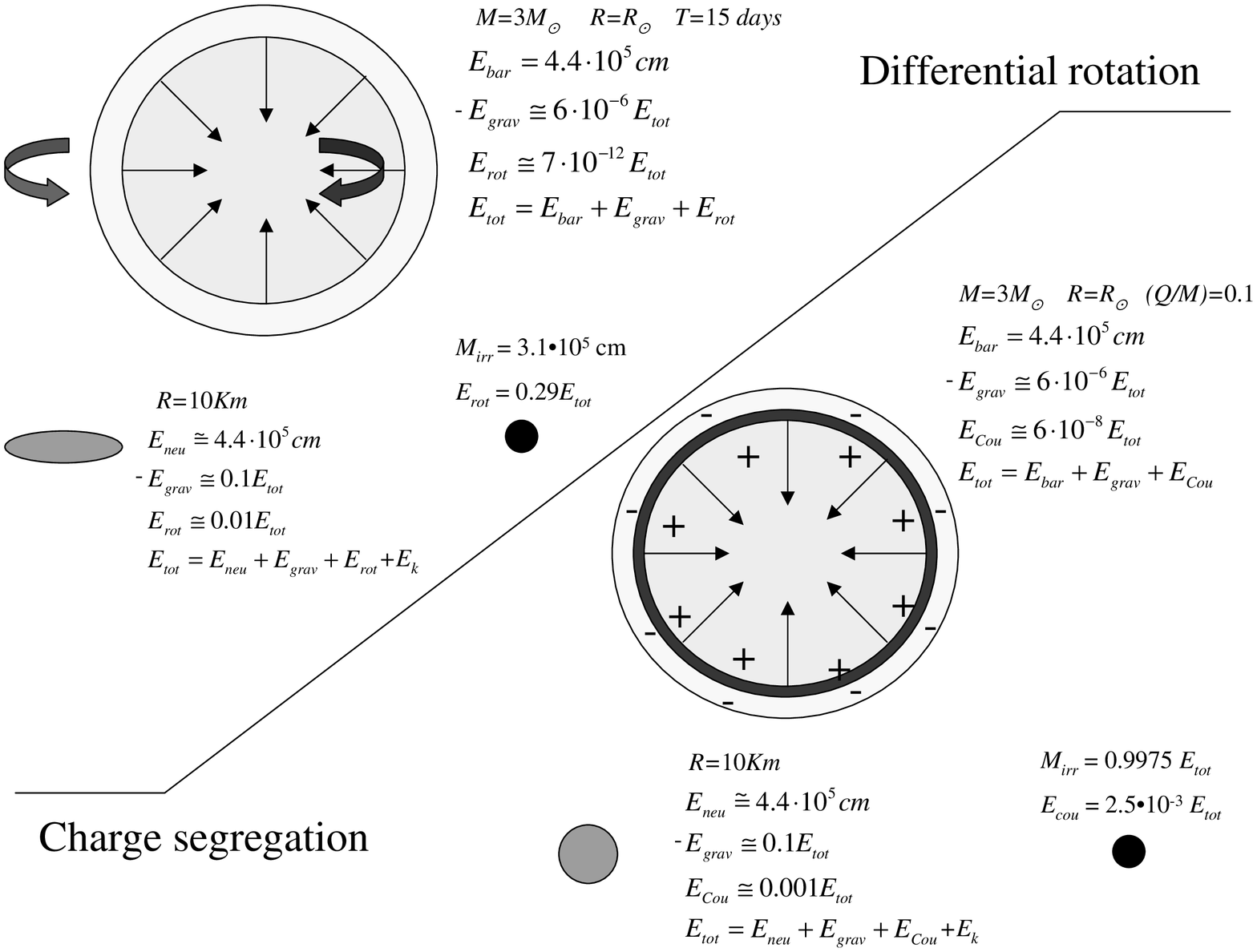}}
\end{center}
\vspace{-0.2cm}
\caption[]{Comparing and contrasting gravitational collapse to a neutron star and to a black hole for a star core endowed with rotation or electromagnetic structure.}
\label{twosph}
\end{figure}

The basic new conceptual point is that GRBs are the most violent transient phenomenon occurring in the universe, so to realize the condition for their occurrence one must look at a transient phenomenon and I propose that this occurs during the most transient phenomenon possibly occurring in the life of a star: the moment of the gravitational collapse. The condition for the creation of the supercritical electromagnetic field required in the Damour and Ruffini work has to be achieved uniquely during the process of gravitational collapse which lasts less than $\sim 30$ seconds for a mass of $10 M_{\odot}$ and the relevant part of the process may be as short as $10^{-2}$ or even $10^{-3}$ seconds (see below). It is appropriate to consider a numerical example here (see Fig~\ref{twosph}) where we have compared and contrasted the gravitational collapse of a star in two limiting cases in which its core of $M=3M_{\odot}$ and radius $R=R_{\odot}$ is either endowed with rotation or with electromagnetic structure. The two possible outcomes of the process of gravitational collapse are considered: either a neutron star of radius of $10 {\rm km}$ or a black hole.

In the case of rotation the core has been assumed to have a rotational period of $\sim 15$ days. For such an initial configuration we have:
\begin{equation}
E_{rot}\simeq 7\times 10^{-12} E_{tot} \ll \left|E_{grav}\right|\simeq 6\times 10^{-6} E_{tot} \ll E_{bar}\simeq 4.4\times 10^5 {\rm cm}\, .
\label{rotcase1}
\end{equation}
In the collapse to a neutron star we have:
\begin{equation}
E_{rot}\simeq 0.01 E_{tot} \ll \left|E_{grav}\right|\simeq 0.1 E_{tot} \ll E_{bar}\simeq 4.4\times 10^5 {\rm cm}\, .
\label{rotcase2}
\end{equation}
The very large increase in the rotational energy is clearly due to the process of gravitational collapse: such a storage of rotational energy is the well known process explaining the pulsar phenomena. The collapse to a black hole has been estimated {\em assuming} the mass-energy formula (see Eqs.(\ref{em})--(\ref{s1})). The overall energetics, for the chosen set of parameters, leads to a solution corresponding to an extreme black hole, for which in principle 29\% of the energy is extractable.

The similar process in the electromagnetic case starts from an initial neutral star with a magnetosphere oppositely charged from a core with
\begin{equation}
\frac{Q}{M\sqrt{G}}=0.1\; .
\label{gc_eq1}
\end{equation}
Let us first evaluate the amount of polarization needed in order to reach the above relativistic condition. Recalling that the charge to mass ratio of a proton is $q_p/\left(m_p\sqrt{G}\right)=1.1\times 10^{18}$, it is enough to have an excess of one quantum of charge every $10^{19}$ nucleons in the core of the collapsing star to obtain such an EMBH after the occurrence of the gravitational collapse. Physically this means that we are dealing with a process of charge segregation between the core and the outer part of the star which has the opposite sign of net charge in order to enforce the overall charge neutrality condition. We here emphasize the name ``charge segregation'' instead of the name ``charge separation'' in order to contrast a very mild charge surplus created in different parts of the star of one electron charge per $10^{19}$ nucleons, keeping the overall charge neutrality, from the much more extreme condition of charge separation in which all the charges of the atomic component of the star are separated.

We then have:
\begin{equation}
E_{cou}\simeq 6\times 10^{-8} E_{tot} \ll \left|E_{grav}\right|\simeq 6\times 10^{-6} E_{tot} \ll E_{bar}\simeq 4.4\times 10^5 {\rm cm}\, .
\label{emcase1}
\end{equation}
In the collapse to the neutron star configuration we have:
\begin{equation}
E_{cou}\simeq 0.001 E_{tot} \ll \left|E_{grav}\right|\simeq 0.1 E_{tot} \ll E_{bar}\simeq 4.4\times 10^5 {\rm cm}\, .
\label{emcase2}
\end{equation}
Once again, the amplification of the electromagnetic energy is due to the process of gravitational collapse. Again, {\em assuming} Eqs.(\ref{em})--(\ref{s1}), the collapse to a black hole for the chosen set of parameters leads to:
\begin{equation}
M_{irr}=0.9975E_{tot}\, , \quad E_{cou}=2.5\times 10^{-3} E_{tot}\, .
\label{emcase3}
\end{equation}
It is during such a process of gravitational collapse to an EMBH black hole that the overcritical field is reached.

The process of charge segregation between the inner core and the oppositely charged outer shell is likely due to the combined effects of rotation and magnetic fields in the earliest phases of the gravitational collapse of the progenitor star.

It is interesting that the two numerical examples given in Fig.~\ref{twosph} point to an underlying theorem: A necessary condition for the occurrence of gravitational collapse as well as for the existence of a stellar equilibrium configuration or of an EMBH is that the numerical value of the gravitational energy be larger than the electromagnetic energy and/or the rotational energy.

In conclusion, the general collapse process involves three distinct moments:
\begin{enumerate}
\item The charge segregation in the progenitor star in the earliest phases of the gravitational collapse. We expect that this process of charge segregation occurs for $r_{ds}\le r \le R_{\odot}$, where $r_{ds}$ is the ``dyadosphere'' radius where the process of vacuum polarization occurs (see next section).
\item The process of amplification of the electric field due to the gravitational collapse leading to overcritical fields and occurring for $r\le r_{ds}$.
\item The latest phases of gravitational collapse to a Kerr-Newman spacetime leading to complex phenomena of ``gravitationally induced electromagnetic radiation'' (see e.g. M. Johnston et al. 1973\cite{ja73}) and of ``electromagnetically induced gravitational radiation'' (see e.g. M. Johnston et al. 1974\cite{ja74}) will tend to reduce both the eccentricity and the angular velocity of the collapsing core.
\end{enumerate}

In this very large theoretical program we have decided to invest in the ``core'' process leading to the occurrence of GRBs. In this sense in our analysis we have focused on point 2 by analyzing the interplay of the amplification of the electromagnetic field and the gravitational field during the process of gravitational collapse in the most simple case, namely in the absence of rotation and of the reduction of the metric given in Eq.~(\ref{KNmetric}) from the case of a Kerr-Newman geometry to the case of a Reissner-Nordstr\"{o}m geometry. This enormously simplifies the computation dealing with spherical symmetry instead of axial symmetry without losing any aspect of the crucial underlying physical process. We shall return to the general case of the Kerr-Newman solution if required by GRB observational evidence.

Therefore the explanation of GRBs in our EMBH model is related to the most transient phenomenon occurring in the life of a star: the process of gravitational collapse.

\section{The ergosphere versus the ``dyadosphere'' of a black hole}

The fundamental new points we have found re-examining the previous work of Damour and Ruffini (1975) \cite{dr75} can be simply summarized:

\begin{itemize}
\item The vacuum polarization process can  occur in an extended region around the black hole called the ``dyadosphere'', extending from the horizon radius $r_+$ out to the ``dyadosphere'' radius $r_{ds}$. Only black holes with a mass larger than the upper limit of a neutron star and up to a maximum mass of $7.2\cdot 10^{6}M_\odot$ can have a ``dyadosphere''.
\item The efficiency of transforming the mass-energy of a black hole into particle-antiparticle pairs outside the horizon can approach 100\%, for black holes in the above mass range. 
\item The created pairs are mainly positron-electron pairs and their number is much larger than the quantity $Q/e$ one would have naively expected on the grounds of qualitative considerations. It is actually given by $N_{\rm pairs}\sim{Q\over e}{r_{ds}\over \hbar/mc}$, where $m$ and $e$ are respectively  the electron mass and charge.  The energy of the pairs and consequently the  emission of the associated electromagnetic radiation  as a function of the black hole mass peaks in the gamma X-ray region.
\end{itemize}

\begin{figure}
\vspace{-.5cm}
\epsfxsize=8.0cm
\begin{center}
\mbox{\epsfbox{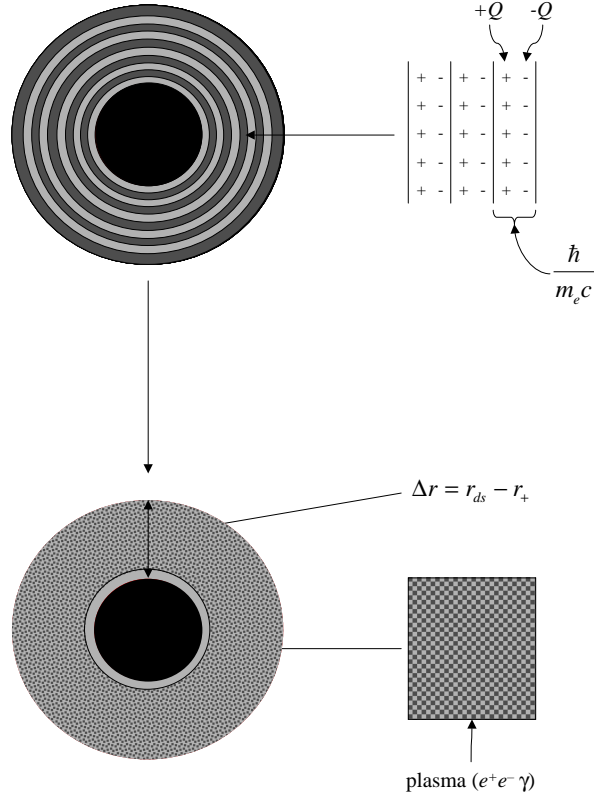}}
\end{center}
\vspace{-0.2cm}
\caption[]{Qualitative features of the vacuum polarization in the ``dyadosphere''. The ``dyadosphere'' can be envisioned as a sequence of concentric capacitors each of thickness $\hbar/\left(m_ec\right)$. After a relaxation time the system thermalizes to an $e^+e^-\gamma$ plasma.}
\label{dyaon}
\end{figure}

Let us now recall the new concept of the ``dyadosphere'' of an EMBH (named for the Greek word {\it dyad\/} for pair) obtained in Preparata, Ruffini and Xue (1998a,b)\cite{prx98ab}. The ``dyadosphere'' for a nonrotating Reissner-Nordstr\"{o}m EMBH (see Fig.~\ref{dyaon}) then lies between the radius
\begin{equation}
r_{\rm ds}=\left({\hbar\over mc}\right)^{1\over2} \left({GM\over c^2}\right)^{1\over2} \left({m_{\rm p}\over m}\right)^{1\over2} \left({e\over q_{\rm p}}\right)^{1\over2} \left({Q\over\sqrt{G} M}\right)^{1\over2}
\label{rc}
\end{equation} 
and the horizon radius 
\begin{equation}
r_{+}={GM\over c^2}\left[1+\sqrt{1-{Q^2\over GM^2}}\right].
\label{r+}
\end{equation}
The number density of pairs created in the ``dyadosphere'' is 
\begin{equation}
N_{e^+e^-}\simeq {Q-Q_c\over e}\left[1+{(r_{ds}-r_+)\over {\hbar\over mc}}\right] \ ,
\label{n}
\end{equation}
where $Q_c=4\pi r_+^2{m^2c^3\over \hbar e}$. The total energy of pairs, converted from the static electric energy, deposited within the ``dyadosphere'' is then
\begin{equation}
E^{\rm tot}_{e^+e^-}={1\over2}{Q^2\over r_+}\left(1-{r_+\over r_{\rm ds}}\right)\left(1-\left({r_+\over r_{\rm ds}}\right)^2\right) \,.
\label{tee}
\end{equation}

The analogies between the ergosphere and the ``dyadosphere'' are many and extremely attractive:

\begin{itemize}
\item Both of them are extended regions around the black hole.
\item In both regions the energy of the black hole can be extracted, approaching the limiting case of reversibility as from Christodoulou and Ruffini (1971).\cite{ruffc}
\item The electromagnetic energy extraction by the pair creation process in the ``dyadosphere'' is much simpler and less contrived than the corresponding process of rotational energy extraction from the ergosphere.
\end{itemize}

\section{The EM pulse of an atomic explosion versus the PEM pulse of a black hole}

The analysis of the radially resolved evolution of the energy deposited within the $e^+e^-$-pair and photon plasma fluid created in the ``dyadosphere'' of an EMBH is much more complex then we had initially anticipated. The collaboration with Jim Wilson and his group at Livermore Radiation Laboratory has been very important for us. We decided to join forces and propose a new collaboration with the Livermore group renewing the successful collaboration with Jim in 1974 (Ruffini and Wilson 1975)\cite{rw75}. We proceeded in parallel: in Rome with simple almost analytic models to be then validated by the Livermore codes (Wilson, Salmonson and Mathews 1997,1998)\cite{wsm97,wsm98}.

For the evolution we assumed the relativistic hydrodynamic equations, for details see Ruffini et al. (1998,1999)\cite{rswx98,rswx99}. We assumed the plasma fluid of $e^+e^-$-pairs, photons and baryons to be a simple perfect fluid in the curved space-time. The baryon-number and energy-momentum conservation laws are 
\begin{eqnarray}
(n_B U^\mu)_{;\mu}&=&(n_BU^t)_{,t}+{1\over r^2}(r^2 n_BU^r)_{,r}= 0\,, \label{contin}\\
(T_\mu^\sigma)_{;\sigma}&=&0\,,
\label{contine}
\end{eqnarray}
and the rate equation: 
\begin{equation}
(n_{e^\pm}U^\mu)_{;\mu}=\overline{\sigma v} \left[n_{e^-}(T)n_{e^+}(T) - n_{e^-}n_{e^+}\right] \,,
\label{econtin}
\end{equation}
where $U^\mu$ is the four-velocity of the plasma fluid, $n_B$ the proper baryon-number density, $n_{e^\pm}$ are the proper densities of electrons and positrons ($e^\pm$), $\sigma$ is the mean pair annihilation-creation cross-section, $v$ is the thermal velocity of the $e^\pm$, and $n_{e^\pm}(T)$ are the proper number-densities of the $e^\pm$ at an appropriate equilibrium temperature $T$. The calculations are continued until the plasma fluid expands, cools and the $e^+e^-$ pairs recombine and the system becomes optically thin.

The results of the Livermore computer code were compared and contrasted with three almost analytical models: (i) spherical model: the radial component of the four-velocity is of the form $U(r)=U{r\over {\cal R}}$, where $U$ is the four-velocity at the surface ($r={\cal R}$) of the plasma, similar to a portion of a Friedmann model, (ii) slab 1: $U(r)=U_r={\rm const.}$, an expanding slab  with constant width ${\cal D}= R_\circ$ in the coordinate frame in which the plasma is moving, (iii) slab 2: an expanding slab with constant width $R_2-R_1=R_\circ$ in the comoving frame of the plasma. We computed the relativistic Lorentz gamma factor $\gamma$ of the expanding $e^+e^-$ pair and photon plasma.

\begin{figure}
\vspace{-.5cm}
\epsfxsize=\hsize
\begin{center}
\mbox{\epsfbox{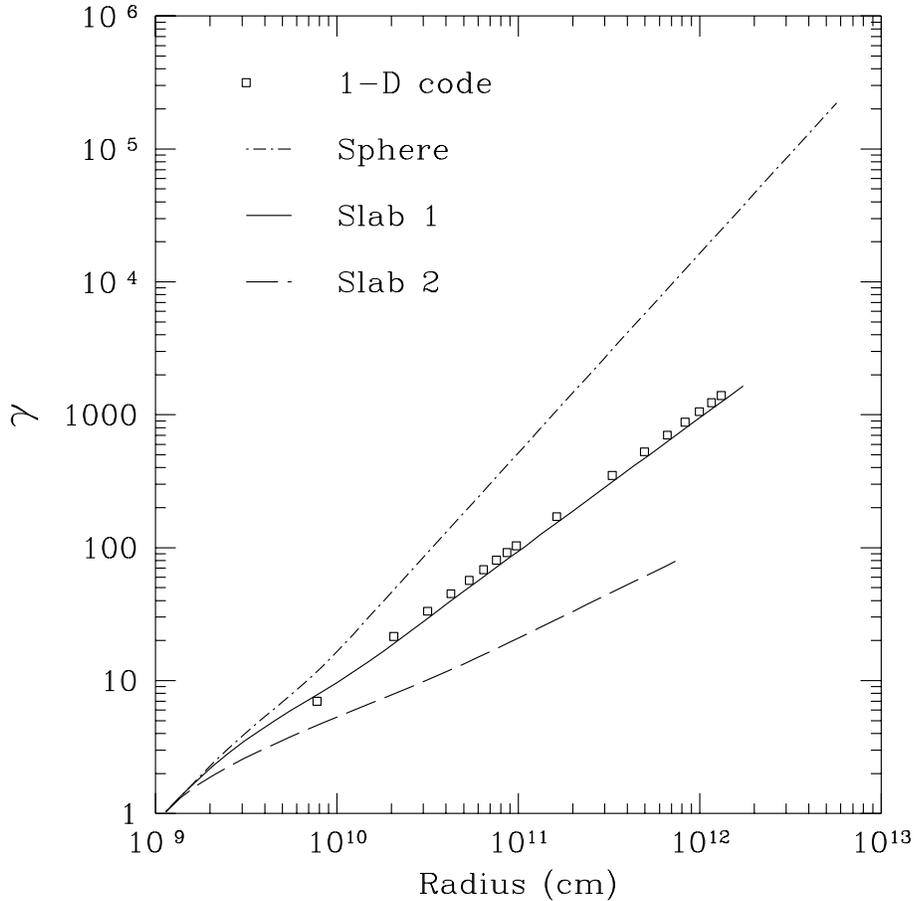}}
\end{center}
\vspace{-0.2cm}
\caption[]{Lorentz $\gamma$ as a function of radius. Three models for the expansion pattern of the $e^+e^-$ pair plasma are compared with the results of the one dimensional hydrodynamic code for a $1000 M_\odot$ black hole with charge $Q = 0.1 Q_{max}$.  The 1-D code has an expansion pattern that strongly resembles that of a shell with constant coordinate thickness. Reproduced from Ruffini, et al. (1999)\cite{rswx99}.}
\label{pic2}
\end{figure}

Figure (\ref{pic2}) shows a comparison of the Lorentz factor of the expanding fluid as a function of radius for all the models. One sees that the one-dimensional code (only a few significant points are plotted) matches the expansion pattern of a shell of constant coordinate thickness. In analogy with the notorious electromagnetic radiation EM pulse of certain explosive events, we called this relativistic counterpart of an expanding pair electromagnetic radiation shell a PEM pulse. Already these preliminary figures show the extraordinary features of this PEM pulse and its dynamical behavior: in the laboratory frame it develops over a scale of $10^{12}\,{\rm cm}$  an increase of the Lorentz gamma factor from $\gamma=1$ to $\gamma\simeq 10^3$.

\begin{figure}
\vspace{-.5cm}
\epsfxsize=\hsize
\begin{center}
\mbox{\epsfbox{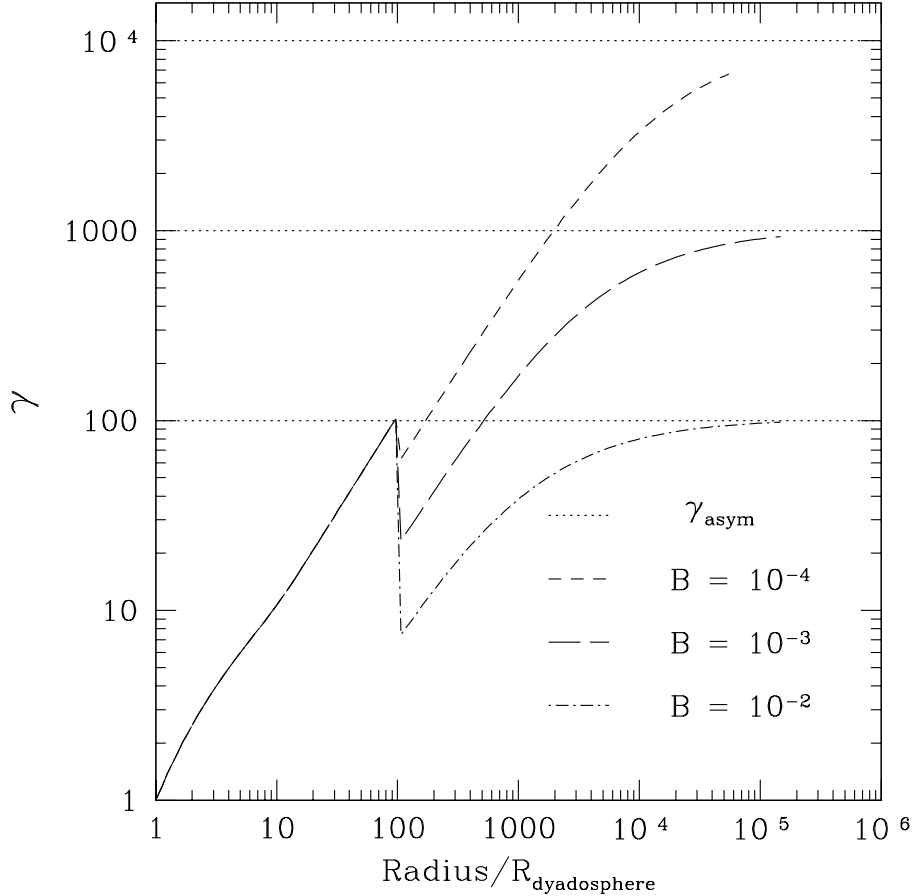}}
\end{center}
\vspace{-0.2cm}
\caption[]{Lorentz gamma factor $\gamma$ as a function of radius for the PEM pulse interacting with the baryonic matter of the remnant (PEMB pulse) for selected values of the baryonic matter. Reproduced from Ruffini, et al. (2000).\cite{rswx00}}
\label{gammab}
\end{figure}

\begin{figure}
\vspace{-.5cm}
\epsfxsize=6.0cm
\begin{center}
\mbox{\epsfbox{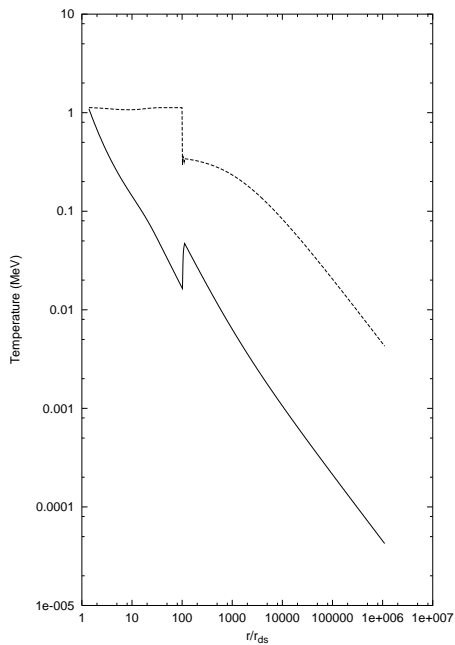}}
\end{center}
\vspace{-0.2cm}
\caption[]{The temperature of the electromagnetic energy component of the PEM pulse, for $r < 100 r_{ds}$ and of the PEMB pulse for $r > 100 r_{ds}$ are given in the laboratory frame (dotted line) and in the comoving frame (solid line). Details may be found in Ref.~\citen{rswx00}.}
\label{f12}
\end{figure}

\begin{figure}
\vspace{-.5cm}
\epsfxsize=6.0cm
\begin{center}
\mbox{\epsfbox{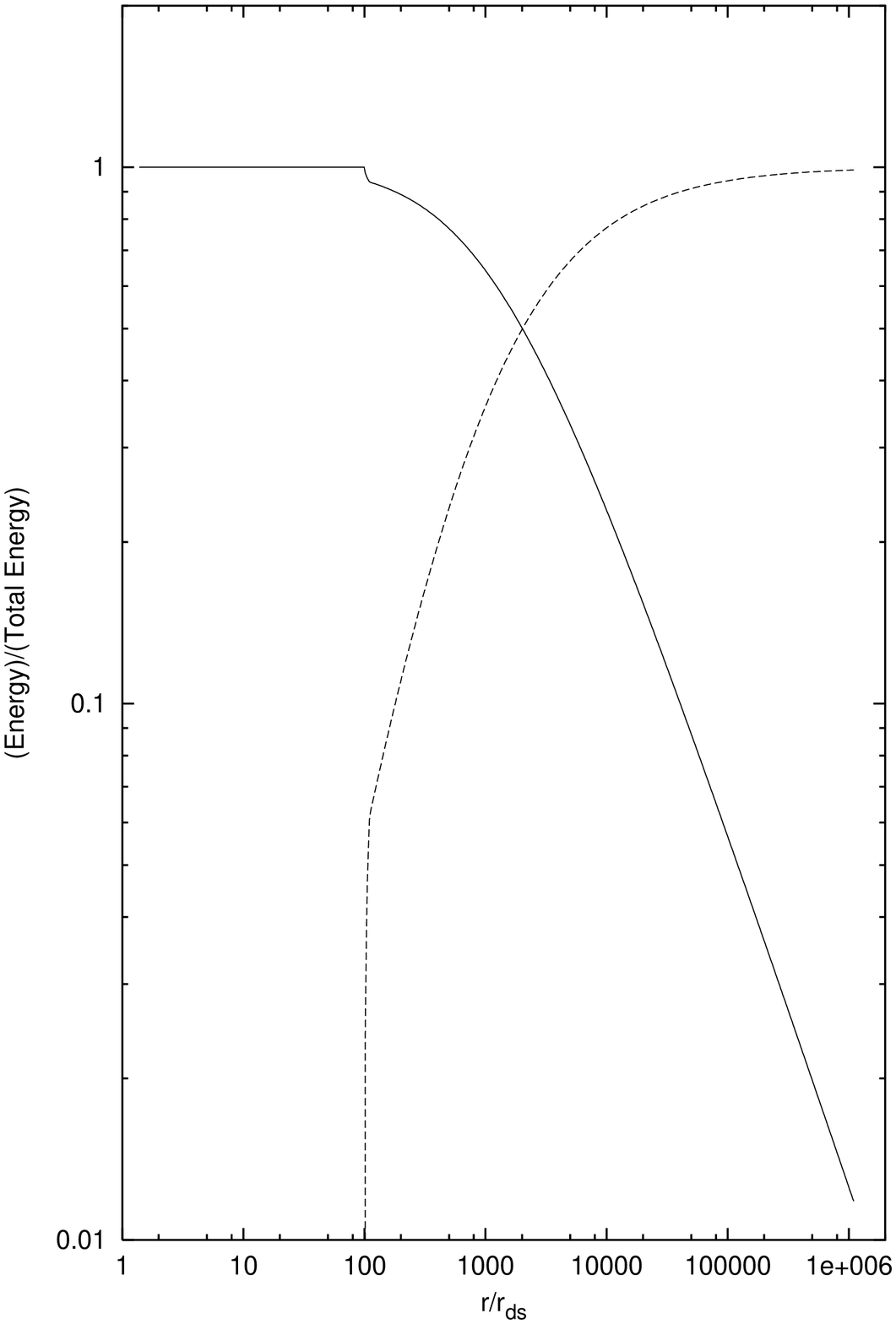}}
\end{center}
\vspace{-0.2cm}
\caption[]{The pair and electromagnetic energy for the PEM pulse $\left(r < 100 r_{ds}\right)$ and PEMB pulse $\left(r > 100 r_{ds}\right)$ measured in the laboratory frame are given as a function of the radial coordinate in units of the ``dyadosphere'' radius (solid line). The kinetic energy of the baryonic matter is also given, for $r > 100 r_{ds}$ (dashed line). Details may be found in Ref.~\citen{rswx00}.}
\label{f13}
\end{figure}

We have also computed the collision of the PEM pulse with the baryonic matter left over in the process of gravitational collapse of the progenitor star. The computation have been carried out using the dimensionless ratio
\begin{equation}
B=\frac{M_Bc^2}{E_{dya}}\, .
\label{Bdef}
\end{equation}
In the collision, the baryonic matter is engulfed by the PEM pulse. A new pulse now originates formed of electron-positron pairs, electromagnetic radiation and baryons: the PEMB pulse. Such a PEMB pulse expands further to values of the Lorentz gamma factor which can be as high as $\gamma\sim 10^4$ (see Fig.~\ref{gammab}).
We have followed the development of the relevant thermodynamical quantities of the PEM and the PEMB pulse during their evolution, both in the comoving and in the laboratory frames (see Fig.~\ref{f12}).
We have also followed the partition of the initial total energy of the ``dyadosphere'' into the pair-electromagnetic energy of the PEM and PEBM pulses and the kinetic energy of the baryons (see Fig.~\ref{f13}).

\begin{figure}
\vspace{-.5cm}
\epsfxsize=\hsize
\begin{center}
\mbox{\epsfbox{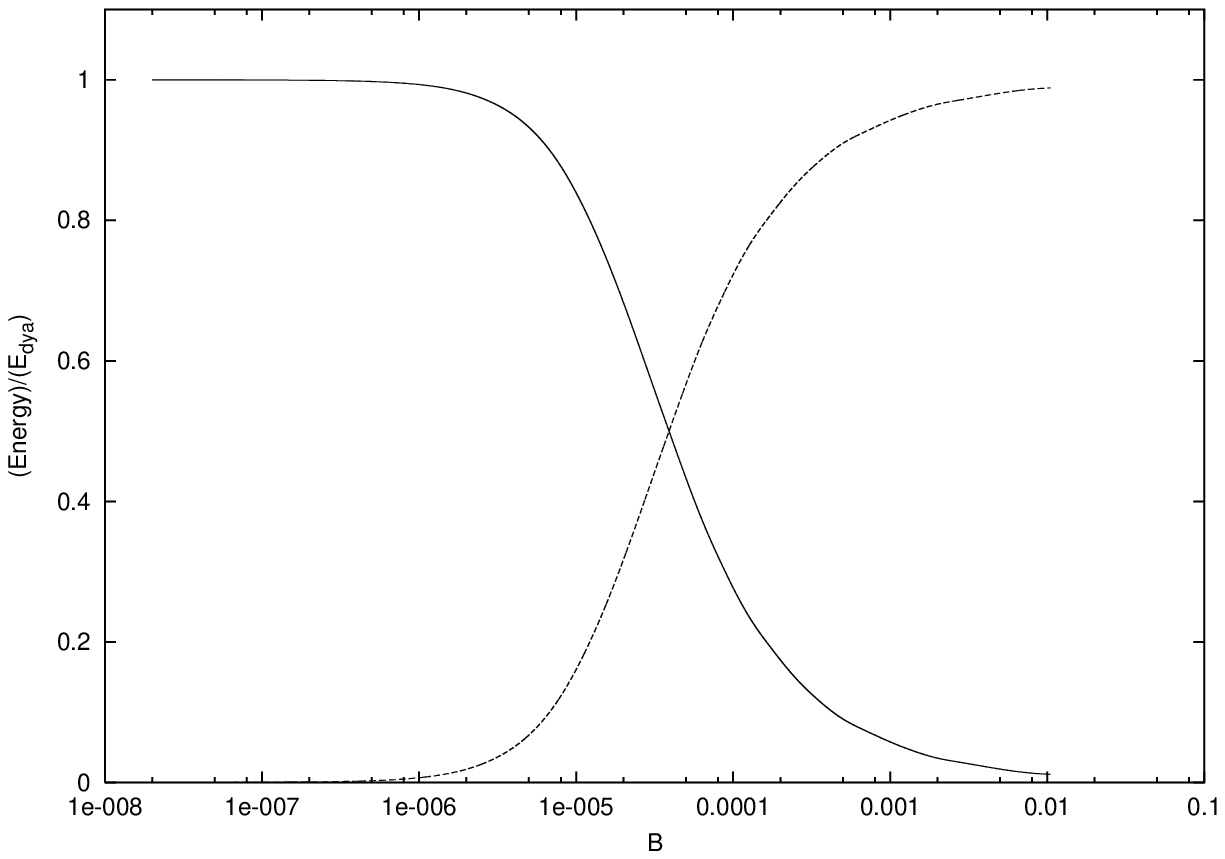}}
\end{center}
\vspace{-0.2cm}
\caption[]{The energy of the P-GRB in units of the ``dyadosphere'' energy is given as a function of the parameter $B$ (solid line). The dotted line represents the kinetic energy of the baryonic component at the transparency point. Details may be found in Ref.~\citen{rswx00}.}
\label{f11}
\end{figure}

The expansion of the PEMB pulse stops as soon as the pulse stops being optically thick and the transparency condition is reached. At this point the proper-GRB (P-GRB) is emitted.\cite{brx01}
Finally we have  plotted as a function of the parameter $B$ both the kinetic energy of the baryonic matter component and the electromagnetic energy of the P-GRB at the moment of transparency (see Fig.~\ref{f11}).

We then proceeded to develop the basic work to describe the afterglow of GRBs (see e.g. Ruffini, et al. 2001)\cite{zzz01}. These results of our theoretical model have reached the point where they can be subjected to a direct comparison with the observational data. We have in fact made the first contact between our theoretical work and the P-GRB observational features in Bianco et al. 2001\cite{brx01}.

\section{Conclusions}

GRBs offer an unprecedented opportunity to probe entire new domains of physics and astrophysics, ranging from high energy particles to the fundamental physical laws and field theories in the extreme spacetimes of black holes and for the first time to give evidence of an energy extraction process for black holes.

\begin{itemize}
\item GRBs are giving the first evidence for the vacuum polarization process in a strong electromagnetic field studied by Sauter-Heisenberg-Schwinger which for many years has been searched for without success in Earth bound accelerators (see e.g. \cite{ga96,la97,la98,ha98}). In order to familiarize the larger scientific community of particle physicists with the extraordinary aspects of this EMBH model for GRBs, we have suggested an analogy with a high-energy accelerator. All the processes described in the previous sections can be visualized as the injector-accelerator phase of an enormous accelerator. The injector phase is characterized by the vacuum polarization process leading to the electron-positron pair creation. The accelerator phase corresponds to the PEM and PEMB pulse phases. Unlike the accelerators on the Earth (CERN, Fermilab, Dubna, etc.), where the acceleration process is due to electromagnetic field, in the cosmic GRB accelerators the acceleration originates in the positron-electron annihilation process and in the mean free path of the photons in such a plasma increasing during the optically thick PEM and PEMB pulse phases. The baryons are carried along by this $e^+e^-$ and photon plasma. This first phase, the injector-accelerator one, terminates with the reaching of the condition of transparency at which point the P-GRB and an accelerated baryonic matter (ABM) pulse are emitted. Clearly both the Lorentz gamma factor of this ABM pulse and the flux are larger then the ones already reached in Earth bound accelerators. If we compare with Fig.~\ref{gf1} where the Lorentz gamma factors of protons at CERN are given, we realize that the Lorentz gamma factor of baryonic matter accelerated by GRBs (see Fig.~\ref{gammab}) can be larger then the ones achievable in the forthcoming years in the Earth bound accelerators. But the enormity of the GRBs accelerators compared to the Earth bound ones is in the differences in fluxes, which can be $10^{38}$ times larger in the astrophysical setting than the ones on the surface of the Earth. The target of the P-GRB and ABM pulse generated in the injector-accelerator phases is represented by the interstellar medium (ISM) and the measuring devices are, in the astrophysical setting, at billions light years of distance, on the surface of the planet Earth. It is no surprise that a wealth of fundamental observations also in particle physics will be obtained from GRBs, in due course.
\begin{figure}
\vspace{-.5cm}
\epsfxsize=\hsize
\begin{center}
\mbox{\epsfbox{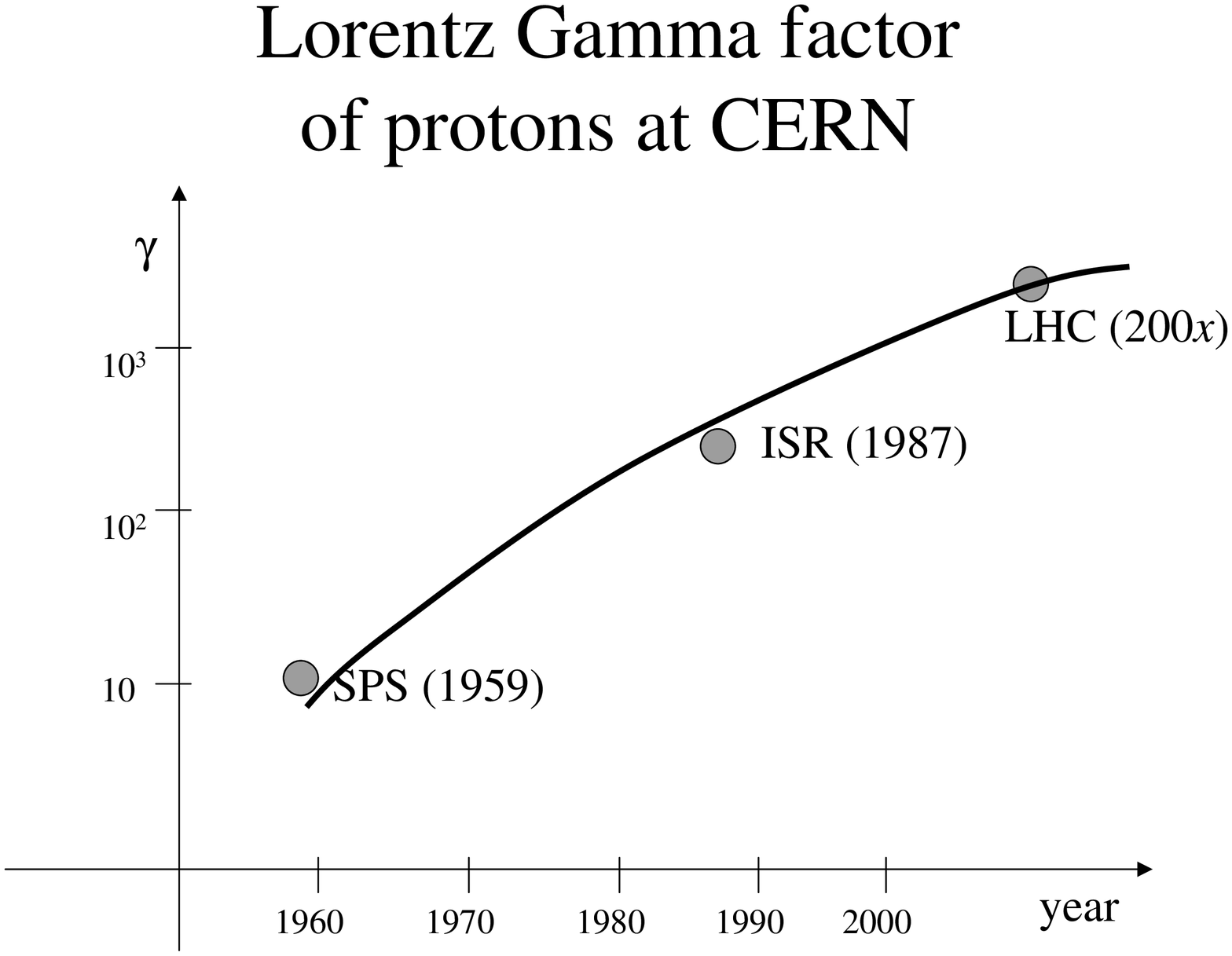}}
\end{center}
\vspace{-0.2cm}
\caption[]{Lorentz gamma factor at CERN over the years.}
\label{gf1}
\end{figure}
\item All the computations on the ``dyadosphere'' we have carried out until now refer to an already formed Reissner-Nordstr\"om black hole. What we need to develop in the near future is the theoretical framework for the time varying process of the dynamical formation of the ``dyadosphere'' and the gradual approach to the horizon (see Fig.~\ref{dyaform}).
\begin{figure}
\vspace{-.5cm}
\epsfxsize=\hsize
\begin{center}
\mbox{\epsfbox{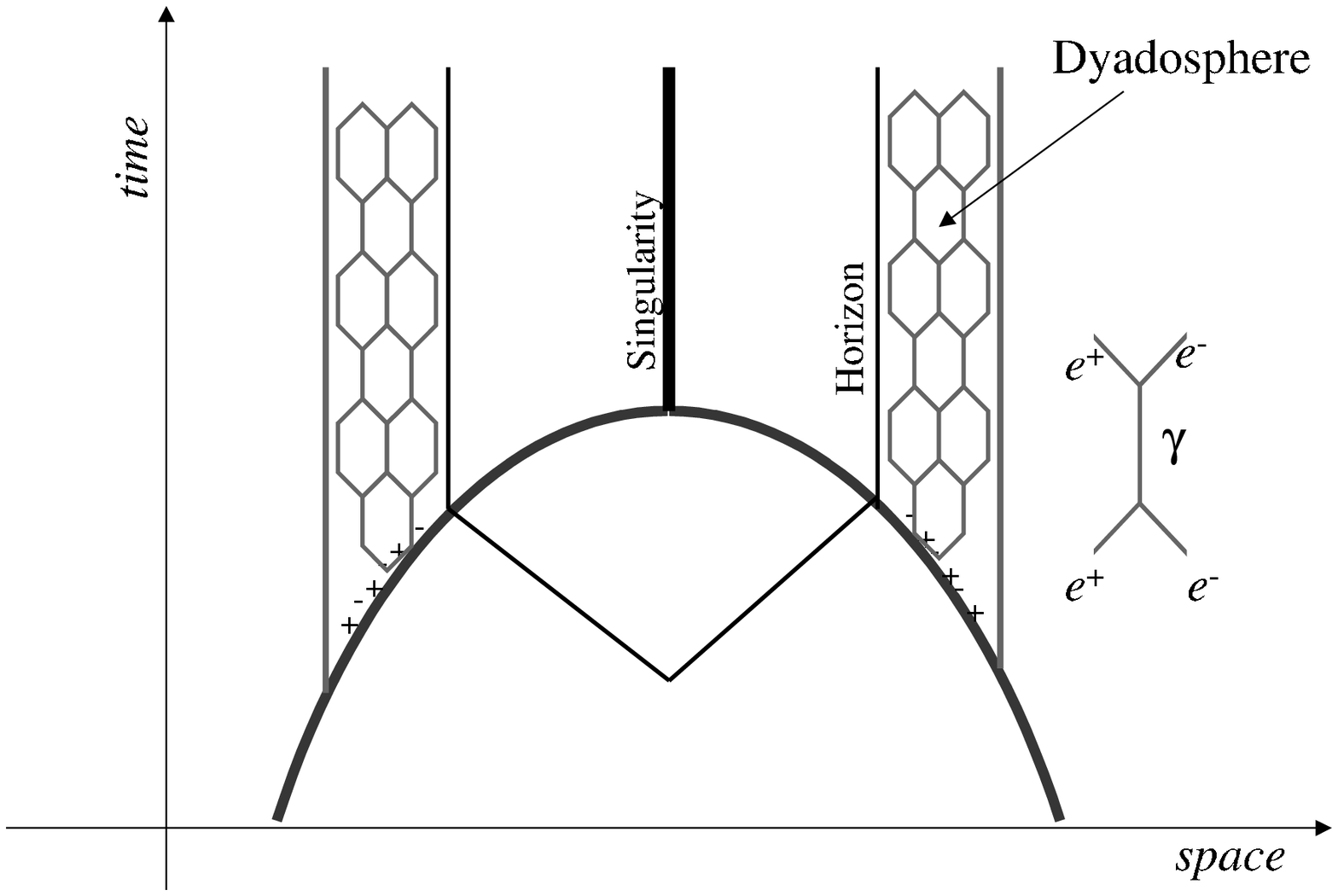}}
\end{center}
\vspace{-0.2cm}
\caption[]{Spacetime diagram of the collapse process leading to the formation of the ``dyadosphere''. As the collapsing core crosses the ``dyadosphere'' radius the pair creation process starts, and the pairs thermalize into a neutral plasma configuration. Then the horizon is  also crossed and the singularity is formed.}
\label{dyaform}
\end{figure}
This time varying evolution can be followed by a sequence of PEM/PEMB pulses emitted outside the collapsing core as the radius of the ``dyadosphere'' $r_{ds}$ is crossed and the horizon radius $r_+$ is approached. From a preliminary analysis we have evidence that there exists a minimum radius $r_{min}>r_+$ such that for $r_+<r<r_{min}$ the PEM/PEMB pulses will be captured by the black hole, while for $r>r_{min}$ the PEM/PEMB pulses will propagate out and will carry their information outwardly, encoded in the structure of the P-GRB. The analysis of the  P-GRB is therefore an essential tool for retracing all the general relativistic effects, including the gravitational redshift and all the time dilation effects which occur in the approach to the horizon of the EMBH.
\item There are enormous differences in the energetics, in the spectra and in the time scale of the radiation processes expected from black holes due to the Damour-Ruffini process and the Beckenstein-Hawking process. In addition to these enormous observational differences, there are also additional conceptual differences between these two processes. While the role of the horizon appear to be predominant in the Beckenstein-Hawking process, it is clear from the computation of the energy extraction process from the ``dyadosphere'' that in the Damour-Ruffini process the role of the horizon in marginal and the energy emission occurs in an extended region between the horizon $r_+$ and the radius of the ``dyadosphere'' $r_{ds}$ (see Fig.~\ref{dyaon}). In addition, the energetics of the energy extraction from black holes is essentially made on the basis of the Christodoulou-Ruffini mass-energy formula (see Eqs.(\ref{em}, \ref{sa}, \ref{s1})). There are a variety of issues still to be addressed. Among these:
\begin{enumerate}
\item What is the physical nature of the irreducible mass and of the Coulomb term in the mass-energy formula?
\item Why 50\% efficiency in the energy extraction process can be reached in an EMBH under the condition of total reversibility\cite{ruffc}?
\item How the reversibility condition in the Damour-Ruffini process, which occurs outside the horizon and in the entire ``dyadosphere'' region, differs from the reversibility condition considered by Christodoulou and Ruffini in the energy extraction, which is instead essentially related to the properties of horizon?
\end{enumerate}
In order to answer these basic physics questions related to the energetics of black holes we need to find a ``gedanken'' process which allows us to describe continuously the transition from a collapsing core to the final asymptotic formation of the black hole. In this sense it appears particularly attractive to use the mathematical solution of the Einstein-Maxwell equations obtained by Werner Israel and his collaborators\cite{i66,dci67} for a charged shell collapsing either on itself or on an already formed EMBH. It seems very probable that, just like the work of Brandon Carter\cite{bc} on the geodesics of Kerr-Newman black holes has offered the essential mathematical and ``gedanken'' transformation tool which led us to the Christodoulou-Ruffini\cite{ruffc} mass formula of black hole, the Israel collapsing shell treatment appears to be the mathematical ``gedanken'' transformation tool essential in probing the energy extraction process for an EMBH (see Fig.~\ref{horform}).
\begin{figure}
\vspace{-.5cm}
\epsfxsize=\hsize
\begin{center}
\mbox{\epsfbox{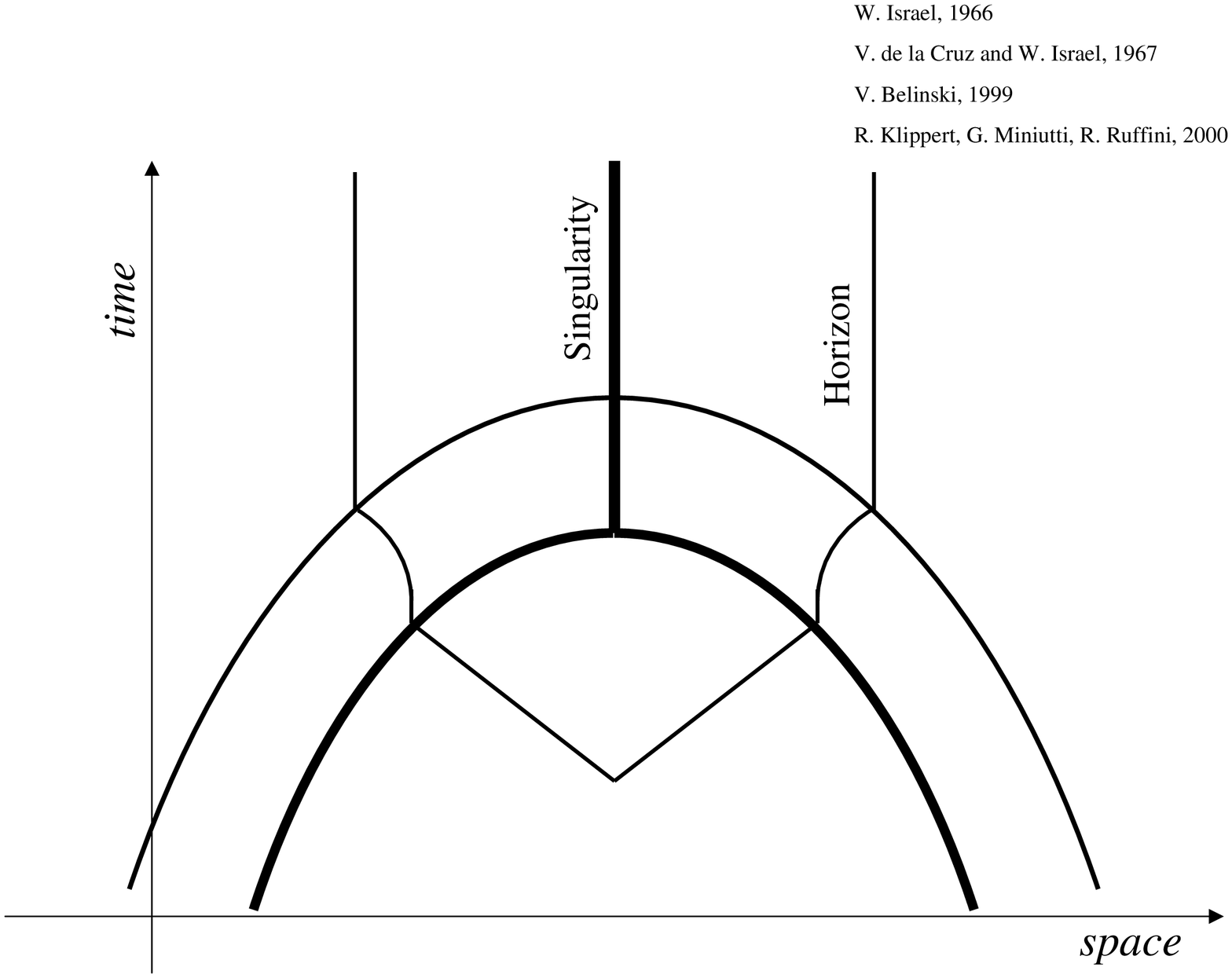}}
\end{center}
\vspace{-0.2cm}
\caption[]{The collapse of two successive charged shells is qualitatively represented. The first one gives rise to an EMBH with a given horizon which is further increased by the accretion of the second shell. This phenomenon can be described exactly with the analytic equations given in Ref.~\citen{crv02}.}
\label{horform}
\end{figure}
The Israel formalism also allows us to evaluate the velocity of the charged shell as it crosses the ``dyadosphere'' radius and finally closes in on the horizon (see Fig.\ref{shelvel}).
\begin{figure}
\vspace{-.5cm}
\epsfxsize=\hsize
\begin{center}
\mbox{\epsfbox{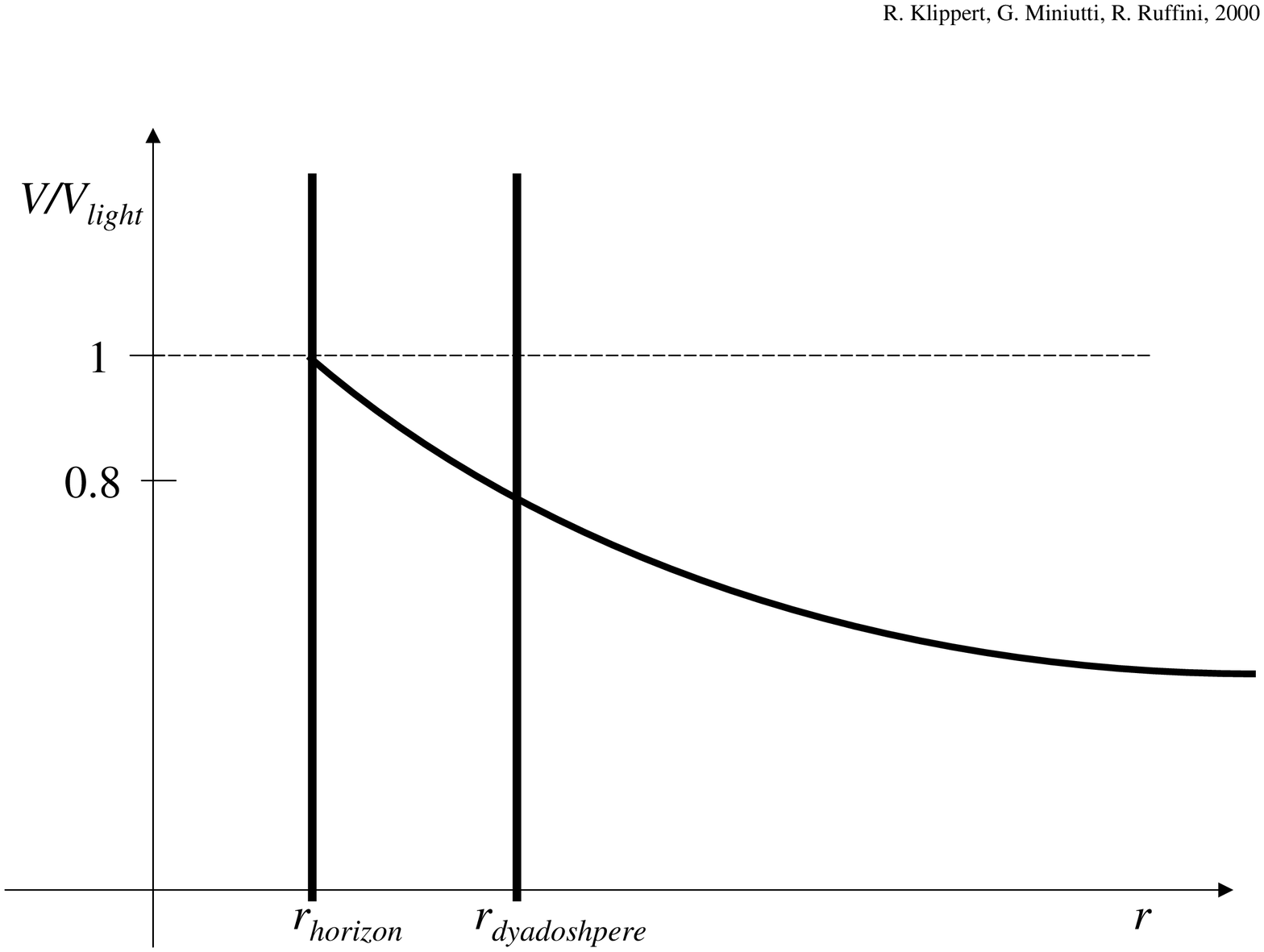}}
\end{center}
\vspace{-0.2cm}
\caption[]{The qualitative behavior of the velocity of a collapsing shell for selected values of the mass of the EMBH and of $\xi$.}
\label{shelvel}
\end{figure}
From the fact that the collapsing core moves inward through the ``dyadosphere'' radius at almost the speed of light, it is then clear that the collapse to a black hole, compared to all the other process of gravitational collapse, is the only one which can guarantee the state of baryonic noncontamination in the ``dyadosphere'' essential to reach the critical value of the electromagnetic field.
\end{itemize}

\section{Some additional results since {\em MGIXMM}}

Since this presentation was made, progress have been made in three major topics:
\begin{enumerate}
\item on the structure of the EMBH and on the physics of the ``dyadosphere'';
\item in establishing some new paradigms in the GRB analysis;
\item in the development of the detailed theoretical model for GRBs and its confrontation with the observations.
\end{enumerate}

We just recall here some major results:\\
{\bf 1.1)} A most outstanding problem confronting mathematical physicists and theoretical physicists in the last 30 years has been the proof of the black hole uniqueness theorem (see Fig.~\ref{tvsetbh}). The technical point is to decouple the equations describing the most general set of electromagnetic and gravitational perturbations of a Kerr-Newman black hole. Such an analysis, once completed, will undoubtly have far reaching consequences, ranging from the microphysical domain of elementary particle physics all the way to the macroscopic phenomena in the black hole domain. In Bini et al.~(2002)\cite{ba02} some progress towards the solution of the analysis of the most general perturbations has been obtained, reducing the equations to the elegant de Rham wave equations for the gravitational and electromagnetic fields in vacuum. A new version of the Teukolksy Master Equation, describing any massless field of spin $s=1/2,1,3/2,2$ in a Kerr black hole, is presented there in the form of a wave equation containing additional curvature terms. These results suggest a relation between curvature perturbation theory in general relativity and the exact wave equations satisfied by the Weyl and the Maxwell tensors, known in the literature as the de Rham-Lichenorowicz Laplacian equations. These Laplacians are discussed both in terms of the Newman-Penrose formalism and in the Geroch-Held-Penrose variant for an arbitrary vacuum spacetime. A perturbative expansion of these wave equations results in a recursive scheme valid for higher orders. This approach, apart from the obvious implications for the gravitational and electromagnetic wave propagation on a curved spacetime, explains and extends the perturbative analysis results in the literature by clarifying their origins in the exact theory.\\
{\bf 1.2)} Turning now from this general scenario to a more detailed analysis of a Reissner-Nordstr\"{o}m geometry, some preliminary necessary steps have been accomplished. In Cherubini et al. (2002)\cite{crv02} we have considered the gravitational collapse of a charged spherical shell with selected boundary conditions: either starting from infinite distance with zero or nonzero kinetic energy, or imploding from a finite distance initially at rest. A new analytic solution has been obtained for such a boundary condition, corresponding both to a collapse into an already formed EMBH or to a collapse in Minkowski space. In both cases we have followed the process of gravitational collapse all the way to the self-closure of the shell by the formation of a horizon.\\
{\bf 1.3)} Using this analytic solution obtained in Cherubini et al.(2002)\cite{crv02}, it has been possible to clarify the independent physical components which contribute to the formation of the EMBH irreducible mass (Ruffini and Vitagliano 2002)\cite{rv02a}. Surprisingly, the irreducible mass does not directly depend on the electromagnetic energy of the imploding shell: it is uniquely a function of the initial baryonic mass, of its gravitational energy and of the kinetic energy of the implosion. The electromagnetic energy is stored around the EMBH and can be extracted by two very different processes as a function of the electromagnetic field strength. 
{\bf a)} When the electric field on the collapsing shell is smaller than ${\cal E}_c$, the process of energy extraction occurs in the effective EMBH ergosphere (\cite{dr73,dhr74}) by a sequence of discrete high energy events, with energy up to $10^{21}$--$10^{27}$\,eV. Such sources can be of relevance for the explanation of the ultra high energy cosmic rays (\cite{bcrx02}). 
{\bf b)} When the electric field on the collapsing shell is larger than ${\cal E}_c$, the conditions relevant to the present article are fulfilled. The energy extraction process occurs in the ``dyadosphere'' and a much larger number of electron and positron pairs are created with typical energies of the order of $10$ MeV which are relevant for the GRB process considered in the present paper.

Having so established and clarified the basic conceptual processes of the energetics of the EMBH, we are now ready, using the new analytic solution found in Cherubini et al. (2002)\cite{crv02}, to approach the dynamical process of vacuum polarization occurring during the formation of an EMBH as qualitatively represented in Fig.~\ref{dyaform}. The study of the  dynamical formation of the ``dyadosphere'' as well as of the electron-positron plasma dynamical evolution will lead to the first possibility of directly observing the general relativistic effects near the EMBH horizon.

Turning now to point 2 above, we moved ahead to fit the observational data on the basis of the EMBH model. We used GRB~991216 as a prototype, both for its very high energetics, which we have estimated in the range of $E_{\rm dya}\sim 9.57\times 10^{52}$ ergs, as well as for the superb data obtained by the Chandra and RXTE satellites. In order to understand the GRB phenomenon, we found it necessary to formulate three new paradigms in our novel approach:\\
{\bf 2.1)} The Relative Space-Time Transformation (RSTT) paradigm (see Ruffini, Bianco, Chardonnet, Fraschetti, Xue (2001a)\cite{lett1}). It relates the observed signals of GRBs to their past light cones, defining the events on the worldline of the source essential for the interpretation of the data. Since GRBs present regimes with unprecedently large Lorentz $\gamma$ factors, also sharply varying with time, particular attention must be given to the constitutive equations relating the four time variables: the comoving time, the laboratory time, the arrival time at the detector, properly corrected by cosmological effects. This paradigm is at the very foundation of any possible interpretation of the data of GRBs.\\
{\bf 2.2)} The Interpretation of the Burst Structure (IBS) paradigm (see Ruffini, Bianco, Chardonnet, Fraschetti, Xue (2001b)\cite{lett2}). It also leads to a reconsideration of the relative roles of the afterglow and burst in GRBs by defining two new phases in this complex phenomenon: a) the injector phase, giving rise to the proper-GRB (P-GRB), and b) the beam-target phase, giving rise to the extended afterglow peak emission (E-APE) and to the afterglow. Such differentiation leads to a natural possible explanation of the bimodal distribution of GRBs observed by BATSE: the short bursts and the long bursts. The agreement with the observational data in regions extending from the horizon of the EMBH all the way out to the distant observer confirms the uniqueness of the model.\\
{\bf 2.3)} The Multiple-Collapse Time Sequence (MCTS) paradigm (see Ruffini, Bianco, Chardonnet, Fraschetti, Xue (2001c)\cite{lett3}) introducing the concept of ``induced gravitational collapse'' and ``induced supernova explosion'' by a GRB. Starting from the data from the Chandra satellite on the iron emission lines in the afterglow of GRB~991216, the following sequence of events is shown to be kinematically possible and consistent with the available data: a) the GRB-progenitor star $P_1$ first collapses to an EMBH, b) the proper GRB (P-GRB) and the peak of the afterglow (E-APE) propagate in interstellar space until the impact on a supernova-progenitor star $P_2$ at a distance $\le 2.69\times 10^{17}$\,cm, and they induce the supernova explosion, c) the accelerated baryonic matter (ABM) pulse, originating the afterglow, reaches the supernova remnants $18.5$ hours after the supernova explosion and gives rise to the iron emission lines. Some considerations on the dynamical implementation of the paradigm are presented. The concept of induced supernova explosion introduced here specifically for the GRB-supernova correlation may have more general application in relativistic astrophysics.\\

Finally, moving to the 3rd problem of the definition of the EMBH model and its confrontation with the observations, we just recall:\\
{\bf 3.1)} The complete model concerning ``the structure of the burst and afterglow of gamma ray bursts I: the radial approximation'' (see Ruffini et al. 2002)\cite{rbcfx02_paperI}. We have presented the entire theoretical background which allowed to formulate the three paradigms mentioned in the above paragraphs 2.1, 2.2 and 2.3. We start from the considerations on the ``dyadosphere'' formation. We then review the basic hydrodynamic and rate equations, the equations leading to the relative spacetime transformations as well as the adopted numerical integration techniques. We then illustrate the five fundamental eras of the EMBH theory: the self-acceleration of the $e^+e^-$ pair-electromagnetic plasma (PEM pulse), its interaction with the baryonic remnant of the progenitor star, the further self-acceleration of the $e^+e^-$ pair-electromagnetic radiation and baryon plasma (PEMB pulse). We then study the approach of the PEMB pulse to transparency, the emission of the proper GRB (P-GRB) and its relation to the ``short GRBs''. Particular attention is given to the free parameters of the theory and to the values of the thermodynamical quantities at transparency. Finally the three different regimes of the afterglow are described within the fully radiative and radial approximations: the ultrarelativistic, the relativistic and the nonrelativistic regimes. The best fit of the theory leads to an unequivocal identification of the ``long GRBs'' as extended emission occurring at the afterglow peak (E-APE). The relative intensities, the time separation and the hardness ratio of the P-GRB and the E-APE are used as distinctive observational tests of the EMBH theory and the excellent agreement between our theoretical predictions and the observations are documented. The afterglow power-law indices in the EMBH theory are compared and contrasted with the ones given in the literature, and no beaming process is found for GRB~991216. Finally, some preliminary results relating the observed time variability of the E-APE to the inhomogeneities in the interstellar medium are presented, as well as some general considerations on the EMBH formation. The general conclusions are then presented based on the three fundamental parameters of the EMBH theory: the ``dyadosphere'' energy, the baryonic mass of the remnant, and the density of the interstellar medium. An in depth discussion and comparison of the EMBH theory with alternative theories is presented as well as indications of further developments beyond the radial approximation.\\
{\bf 3.2)} The unvailing of the physical processes which lie at the bases of time variability of GRBs has been developed in Ruffini et al. (2001)\cite{lett5}. In this paper we relate the observed substructure in the peak of the afterglow to the interaction of the ABM pulse with the ISM. It is found that with the exception of the relatively inconspicuous but scientifically very important signal originating from the initial ``proper gamma ray burst'' (P-GRB), all the other spikes and time variabilities can be explained by the interaction of the accelerated-baryonic-matter pulse with inhomogeneities in the interstellar matter. This can be demonstrated by using the RSTT paradigm as well as the IBS paradigm, to trace a typical spike observed in arrival time back to the corresponding one in the laboratory time. Using these paradigms, the identification of the physical nature of the time variablity of the GRBs can be made most convincingly. We make explicit the dependence of a) the intensities of the afterglow, b) the spike amplitude and c) the actual time structure on the Lorentz gamma factor of the accelerated-baryonic-matter pulse. In principle it is possible to read off from the spike structure the detailed density contrast of the interstellar medium in the host galaxy, even at very high redshift.\\
{\bf 3.3)} The proof of the validity of the approximation adopted in 3.1 and 3.2 has finally be presented in ``On the structure of the burst and afterglow of gamma ray bursts II: the angular spreading'' (Ruffini et al. 2002)\cite{rbcfx02_paperII}. Using GRB~991216 as a prototype, the relativistic angular spreading in the computation of the afterglow within the EMBH model is presented. Comparison of the present results with the ones based on the radial approximation confirm the validity of the conclusions reached within that approximation. It is shown that the intensity substructures observed in the extended afterglow peak emission (E-APE) do indeed originate in the collision between the accelerated baryonic matter (ABM) pulse with inhomogeneities in the interstellar medium (ISM) in a regime with Lorentz factor $\gamma \sim 310$. The crossing of ISM inhomogeneities of sizes $\Delta R\sim 10^{15}\, {\rm cm}$ occurs in a detector arrival time interval of $\sim 0.4\, {\rm s}$ implying an apparent superluminal behavior of $\sim 10^5c$. Our results are compared and contrasted with those in the current literature.\\

Finally progress toward the establishment of the theorem quoted in section~\ref{newparadigm} is being accomplished and some preliminary results and astrophysical consequences are appearing in the proceedings volume {\it Fermi and Astrophysics\/}\cite{fa}


\begin{thebibliography}{99}

\bibitem{nccept65}
E.T. Newman, E. Couch, R. Chinnapared, A. Exton, A. Prakash and R. Torrence, 
{\it J.\ Math.\ Phys.} {\bf 6}, 918 (1965).

\bibitem{gr78}
R. Giacconi and R. Ruffini, Eds.\ and coauthors, 
{\it Physics and Astrophysics of Neutron Stars and Black Holes} 
(North Holland, Amsterdam, 1978).

\bibitem{snyder}
J.R. Oppenheimer and H. Snyder, 
{\it Phys.\ Rev.} {\bf 56}, 455 (1939).

\bibitem{s34} 
C. St\o rmer,   
{\it Astrophysica Norvegica} {\bf 1}, 1 (1934).

\bibitem{bc}
B. Carter, 
{\it Phys.\ Rev.} {\bf 174}, 1559 (1968).

\bibitem{rrw}
M. Rees, R. Ruffini and J.A. Wheeler,  
 {\it ``Black Holes, Gravitational Waves and Cosmology''}, 
(Gordon and Breach,  New York, 1974) (also in Russian, MIR 1973).

\bibitem{rw71}
R. Ruffini and J.A. Wheeler, 
{\it Introducing the Black Hole},
{\it Physics Today} {\bf 24} (1), 30 (1971).

\bibitem{ReggeW}
T. Regge and J.A. Wheeler, 
{\it Phys.\ Rev.} {\bf 108}, 1063 (1957).

\bibitem{Zerilli1}
F.J. Zerilli, 
{\it Phys.\ Rev.} {\bf D2}, 2141 (1970). 

\bibitem{Zerilli2}
F.J. Zerilli, 
{\it Phys.\ Rev.} {\bf D9}, 860 (1974). 

\bibitem{teukolsky}
S.A. Teukolsky,
{\it Astrophys.\ J.} {\bf 185}, 635 (1973).

\bibitem{lee}
C. H. Lee, 
{\it J.\ Math.\ Phys.} {\bf 17}, 1226 (1976); 
{\it Prog.\ Theor.\ Phys.} {\bf 66}, 180 (1981).

\bibitem{chandra}
S. Chandrasekhar,
{\it The Mathematical Theory of Black Holes}, 
(Clarendon Press, Oxford, 1983), 
see also S. Chandrasekhar, 
{\it Proc.\ R.\ Soc.\ Lond.} {\bf A349}, 571 (1976).

\bibitem{cru}
G. Cruciani, in {\it Proceedings of the Third Icra Network Workshop (1999)\/},  C. Cherubini and  R. Ruffini, Eds., Editrice Compositori, (Bologna, 2001) and in {\it Nuovo Cim.} {\bf B115}, 687 (2000)

\bibitem{chrr}
C. Cherubini and R. Ruffini, 
in {\it Proceedings of the Third Icra Network Workshop (1999)\/},  C. Cherubini and  R. Ruffini, Eds., Editrice Compositori, (Bologna, 2001) and in {\it Nuovo Cim.} {\bf B115}, 699 (2000).

\bibitem{bcjr1}
D. Bini, C. Cherubini, R.T. Jantzen and R. Ruffini,
Wave equation for tensor valued $p$-forms: application to the Teukolsky master equation,
preprint (2001).

\bibitem{ba02}
D. Bini, C. Cherubini, R.T. Jantzen and R. Ruffini,
{\it Prog.\ Theor.\ Phys.} {\bf 107}, 967 (2002).

\bibitem{p69}
R. Penrose,
{\it Nuovo Cim.\ Rivista} {\bf 1}, 252 (1969).

\bibitem{chris1}
D. Christodoulou, 
{\it Phys.\ Rev.\ Lett.} {\bf 25}, 1596 (1970).

\bibitem{ruffx}
R.  Ruffini and J.A. Wheeler, 
``Relativistic Cosmology from Space Platforms'' 
in {\it Proceedings of the Conference on Space Physics\/}, 
Hardy V. and  Moore H., Eds., E.S.R.O. Paris, (1971). 
The preparation of this report took more then one year and the authors were unwilling to publish parts of it before the final publication. In order to avoid delays,  the results of the energy extraction process from a Kerr black hole, as well as the definition of the ``ergosphere'', were inserted as Fig.~2 in the Christodoulou 1970 paper, published 30 November 1970. 

\bibitem{fr} 
R.M. Floyd and R. Penrose, 
{\it Nature} {\bf 229}, 177 (1971), submitted 16 December 1970.

\bibitem{ruffc}
D. Christodoulou and R. Ruffini,
{\it Phys.\ Rev.} {\bf D4}, 3552 (1971).

\bibitem{dw73}
B. de Witt and C. de Witt, Eds., 
{\it Black Holes}, (Gordon and Breach, New York, 1973).

\bibitem{rr74}
C. Rhoades and R. Ruffini,  
{\it On the maximum mass of neutron stars}, 
{\it Phys.\ Rev.\ Lett.} {\bf 32}, 324 (1974).

\bibitem{lr73}
R.W. Leach and R. Ruffini, 
{\it Astrophys.\ J.} {\bf 180}, L15--L18 (1973).

\bibitem{rufsolv}
R. Ruffini, ``Neutron Stars, Black Holes and Binay X-Ray Sources'', in Proceedings of the Sixteenth Solvay Conference on Physics at the University of Bruxelles, Editions de l'Universit\'e de Bruxelles, 1974, pp. 394--424.

\bibitem{gr75} 
H. Gursky and R. Ruffini, Eds.\ and coauthors, 
{\it Neutron Stars, Black Holes and Binary X-ray Sources} 
(D.~Reidel, Dordrecht, 1975).

\bibitem{dr75}	
T. Damour and R. Ruffini, 
{\it Phys.\ Rev.\ Lett.} {\bf 35}, 463 (1975).

\bibitem{he35}
W. Heisenberg and H. Euler, 
{\it Zeits.\ Phys.} {\bf 98}, 714 (1935).

\bibitem{s51} 
J. Schwinger, 
{\it Phys.\ Rev.} {\bf98}, 714 (1951).

\bibitem{c97}
E. Costa et al., 
{\it Nature} {\bf 387}, 783 (1997).

\bibitem{mgixmm}
{\it Proceedings of the Ninth  Marcel Grossmann Meeting on General Relativity}, 
V. Gurzadyan, R. Jantzen and R. Ruffini, Eds.\ 
(World Scientific, Singapore, 2001).

\bibitem{rw75} 
R. Ruffini and J.R. Wilson, 
{\it Phys.\ Rev.} {\bf D12}, 2959 (1975).

\bibitem{s70}
V.F. Shvartsman, 
{\it Sov.\ Phys.\ JETP} {\bf 33}, 475 (1970).

\bibitem{punsly_book}
B. Punsly,  
{\it Black Hole Gravitohydromagnetics}, Springer, 2001.

\bibitem{gj69}
P. Goldreich and W.H. Julian,
{\it Ap.\ J.} {\bf 157}, 869 (1969).

\bibitem{ja73}
M. Johnston, R. Ruffini and F. Zerilli, 
{\it Phys.\ Rev.\ Lett.} {\bf 31}, 1317 (1973).

\bibitem{ja74}
M. Johnston, R. Ruffini and F. Zerilli, 
{\it Phys.\ Lett.} {\bf 49B}, 185 (1974).

\bibitem{prx98ab}
G. Preparata, R. Ruffini and S.-S Xue,   
submitted to {\it Phys.\ Rev.\ Lett.} and  
{\it Astronomy and Astrophysics} {\bf 338}, L87 (1998). 

\bibitem{wsm97}
J.R. Wilson, J.D. Salmonson and G.J. Mathews,
in {\it Gamma-Ray Bursts: 4th Huntsville Symposium},  
C.A. Meegan, R.D. Preece and T.M. Koshut, Eds.\ (A.I.P., 1997).

\bibitem{wsm98}
J.R. Wilson, J.D. Salmonson, and G.J. Mathews, in 
{\it 2nd Oak Ridge Symposium on Atomic and Nuclear Astrophysics} 
(IOP Publishing, 1998). 

\bibitem{rswx98}
Ruffini, R., Salmonson, J.D., Wilson, J.\ R., S.-S Xue,
{\it Astron.\ Astroph.\ Suppl.\ Ser.} {\bf 138}, 511 (1998).

\bibitem{rswx99}
Ruffini, R., Salmonson, J.D., Wilson, J.\ R., S.-S Xue, 
{\it Astron.\ Astroph.} {\bf 350}, 334  (1999).

\bibitem{rswx00}
Ruffini, R., J.D. Salmonson, J.R. Wilson and S.-S Xue, 
{\it Astron.\ Astroph.}  {\bf359}, 855 (2000).

\bibitem{brx01}
C.L. Bianco, R. Ruffini, S.-S Xue,
{\it Astron.\ Astroph.} {\bf368}, 377 (2001).

\bibitem{zzz01}
R. Ruffini, C.L. Bianco, P. Chardonnet, F. Fraschetti and S.-S. Xue, 
submitted to {\it Astron.\ Astroph.} (2001)d

\bibitem{ga96}
Ganz, R. et al., {\it Phys.\ Lett.\ B} {\bf 389}, 4 (1996).

\bibitem{la97}
Leinberger, U. et al., {\it Phys.\ Lett.\ B} {\bf 394}, 16 (1997).

\bibitem{la98}
Leinberger, U. et al., {\it Eur.\ Phys.\ J.\ A} {\bf 1}, 249 (1998).

\bibitem{ha98}
Heinz, S. et al., {\it Eur.\ Phys.\ J.\ A} {\bf 1}, 27 (1998).

\bibitem{i66}
W. Israel, 
{\it Nuovo Cim.} {\bf 44B}, 1 (1967).

\bibitem{dci67}
V. De la Cruz, W. Israel,
{\it Nuovo Cim.} {\bf 51A}, 744 (1967).

\bibitem{crv02}
C. Cherubini, R. Ruffini and L. Vitagliano,
in preparation (2002).

\bibitem{rv02a}
R. Ruffini and L. Vitagliano, in preparation (2002)a.

\bibitem{dr73}
G. Denardo and R. Ruffini,
{\it Phys.\ Lett.} {\bf 45B}, 259 (1973).

\bibitem{dhr74}
G. Denardo, L. Hively and R. Ruffini,
 {\it Phys.\ Lett.} {\bf 50B}, 270  (1974).

\bibitem{bcrx02}
C.L. Bianco,  P. Chardonnet,  R. Ruffini and S.-S. Xue, 
in preparation (2002). 

\bibitem{lett1}
R. Ruffini, C.L. Bianco, P. Chardonnet, F. Fraschetti, S.-S. Xue,
{\it Ap.\ J.\ Lett.} {\bf 555}, L107 (2001)a.

\bibitem{lett2}
R. Ruffini, C.L. Bianco, P. Chardonnet, F. Fraschetti and S.-S. Xue, 
{\it Ap.\ J.\ Lett.} {\bf 555}, L113 (2001)b.

\bibitem{lett3}
R. Ruffini, C.L. Bianco, P. Chardonnet, F. Fraschetti and S.-S. Xue, 
{\it Ap.\ J.\ Lett.} {\bf 555}, L117 (2001)c. 

\bibitem{rbcfx02_paperI}
R. Ruffini, C.L. Bianco, P. Chardonnet, F. Fraschetti and S.-S. Xue,
submitted to {\it Astron.\ Astroph.} (2002)d. 

\bibitem{lett5}
R. Ruffini, C.L. Bianco, P. Chardonnet, F. Fraschetti and S.-S. Xue, 
{\it Nuovo Cim.} {\bf 116B}, 99 (2001).

\bibitem{rbcfx02_paperII}
R. Ruffini, C.L. Bianco, P. Chardonnet, F. Fraschetti and S.-S. Xue, 
 submitted to {\it Astron.\ Astroph.} (2002)e.

\bibitem{fa}
V. Gurzadyan, R. Ruffini (editors and co-authors), ``Fermi and Astrophysics'', World Scientific (Singapore, in press)

\end{thebibliography}
\end{document}